\documentclass[11pt]{article}
\usepackage{gbmacros}
\usepackage[colorlinks=true,citecolor=blue]{hyperref}
\usepackage{fourier,charter,graphicx,color,array}

\usepackage{courier}

\def\maketiToda lattice{\par\noindent{\Large\bf\thetiToda lattice}\\[1.4ex]
 {\large\theauthor}\\[0.6ex]
 \textit{\thetextinfo}\\[0.6ex]
 {\small\today}\par\vglue1.4\bigskipamount}
\def\tiToda lattice#1{\def\thetiToda lattice{#1}}
\def\author#1{\def\theauthor{#1}}
\def\textinfo#1{\def\thetextinfo{#1}}
\allowdisplaybreaks
\usepackage{tocloft}
\def\fbf#1{\setbox0=\hbox{$#1$}\kern-0.10\wd0
  \lower0.02em\copy0\kern-\wd0 \lower0.02em\hbox{\kern+0.04em\copy0}\kern-\wd0
  \raise0.00em\copy0\kern-\wd0 \raise0.00em\hbox{\kern-0.04em\box0}}
\def\T{{\mathrm{T}}}

\makeatletter
\let\tru@int=\int
\def\int{\mathop{\textstyle\tru@int}\limits}
\def\overl@ss#1#2{\vcenter{\offinterlineskip
        \ialign{$\m@th#1\hfil##\hfil$\crcr#2\crcr<\crcr } }}
\def\overgr@at#1#2{\vcenter{\offinterlineskip
        \ialign{$\m@th#1\hfil##\hfil$\crcr#2\crcr>\crcr } }}
\def\gl{\mathrel{\mathpalette\overl@ss>}}
\def\lg{\mathrel{\mathpalette\overgr@at<}}
\def\d{\mathrm{d}}
\def\Natural{\mathbb{N}}
\def\Integer{\mathbb{Z}}
\def\Real{\mathbb{R}}

\def\pvint{\int\kern-0.94em-\kern0.2em}
\let\^=\hat
\let\==\bar
\let\@=\mathbf

\def\d{\mathrm{d}}
\def\e{\mathrm{e}}

\def\sn{\mathop{\rm sn}\nolimits}
\def\O#1{^{(#1)}}
\def\~#1{\tilde{\mathbf{#1}}}
\let\~=\tilde
\let\l=\lambda
\makeatother

\advance\textwidth -2mm
\advance\hoffset 1mm
\advance\textheight 10mm
\advance\voffset -5mm
\advance\footskip -2mm
\numberwithin{equation}{section}
\def\be{\begin{equation}}
\def\ee{\end{equation}}
\def\bse{\begin{subequations}}
\def\ese{\end{subequations}}
\definecolor{deeppurple}{rgb}{0.6, 0.0, 0.6}
\def\hreftiToda lattice#1#2{\href{#1}{``#2''}}

\tiToda lattice{\color{red}On the Whitham modulation equations for the Toda lattice and the\\[0.4ex]quantitative characterization of its dispersive shocks}
\author{Gino Biondini$^1$, Christopher Chong$^2$ and Panayotis Kevrekidis$^3$}
\textinfo{%
$^1$ Department of Mathematics, State University of New York at Buffalo, Buffalo, New York 14260\\
$^2$ Department of Mathematics, Bowdoin College, Brunswick, Maine 04011\\
$^3$ Department of Mathematics and Statistics, University of Massachusetts, Amherst, Massachusetts 01003
}

\begin{document}
\maketiToda lattice

\kern-\bigskipamount
\begin{quote}
\textbf{Abstract.}
The aim of this work is multifold. Firstly, it
intends to present a complete, quantitative and self-contained description of the periodic traveling wave solutions and 
Whitham modulation equations for the Toda lattice, combining results from different previous works in the literature.
Specifically, we connect the Whitham modulation equations and a detailed expression for the periodic traveling wave solutions 
of the Toda lattice. Along the way, some
key details are filled in, such as %e.g.,
%and also filling in some details in previous works
%(specifically, 
the explicit expression of the characteristic speeds of the genus-one Toda-Whitham system. 
Secondly, we use these tools to obtain a detailed quantitative characterization of the dispersive shocks of the Toda system. 
Lastly, we validate the relevant analysis by performing a detailed comparison with direct numerical simulations. 
\end{quote}

%%%%%%%%%%%%%%%%%%%%%%%%%%%%%%%%%%%%%%%%%%%%%%%%%%%%%%%%%%%%%%%%%%%%%
\section{Introduction}

The study of shock waves in dissipative~\cite{Smoller},
as well as in dispersive~\cite{Whitham74} systems
has a time-honored history. 
While the former are, arguably, more well-known and
the corresponding theory is more well-developed,
the latter in the form of the so-called dispersive
shock waves (DSWs) have seen numerous developments
in the last few decades. 
This is both due to the emergence of theoretical approaches, such 
as Whitham modulation theory~\cite{Whitham74,GP73,Karpman} 
(see also the reviews~\cite{dsw,Mark2016}) for their study,
but also due to numerous experimental developments more recently that have made
such patterns more experimentally accessible.
These include, but are not limited to, nonlinear optical systems~\cite{fleischer,Trillo2018}, 
ultracold Bose-Einstein condensates~\cite{Hoefer2006,davis}, 
as well as fluid conduits~\cite{Hoefer2016,Hoefer2018}.

While the above mentioned developments
(and the vast majority of the literature on shock waves) 
are rooted in the study of continuous systems, 
recent theoretical and experimental developments have showcased
the relevance of exploring and understanding such phenomena also in (spatially) discrete realms. 
Some of the early studies~\cite{ruffo},
focused on the analogy of discrete problems with corresponding continuum ones 
(such as, e.g., of the Fermi-Pasta-Ulam-Tsingou
model with the well-established~\cite{GP73} Korteweg-de Vries (KdV) case). 
Nevertheless, more recently, it has become evident that metamaterial lattices can feature
genuinely discrete and experimentally observable DSW features
\cite{Nester2001,Hascoet2000,Herbold07,Molinari2009,shock_trans_granular,HEC_DSW}
(indeed, a finding that was 
%originally, to our knowledge, 
established at least as early as
the work of~\cite{first_DSW}).
It is relevant to highlight here that in addition to these material lattices,
optical systems of waveguide arrays~\cite{fleischer2} have also been used to experimentally illustrate
such discrete analogues. 
More recently, the study of ultra-slow shock waves 
in a tunable magnetic lattice~\cite{talcohen} has been added to the
list of relevant systems, further diversifying the realm of applications of discrete DSW studies. 

As may naturally be expected, the discrete systems that may feature
DSWs are prototypically ones where features related to
integrability~\cite{AS81,AblowitzPrinariTrubatch} 
(Lax pairs, infinitely many conservation laws, and the ability
to solve the system analytically via inverse scattering) 
are absent.
Nevertheless, the ability to utilize exactly solvable integrable systems 
to demonstrate ideas and develop analytical solutions related to DSWs has had
a time-honored history in continuum systems, as shown in~\cite{Whitham74,GP73,dsw,Mark2016}.
Indeed, similar analogies between discrete non-integrable systems 
(such as the granular lattice \cite{nesterenko07}) and
integrable ones (such as the Toda lattice~\cite{toda1967}) have
proved valuable in connection to other wave structures such as 
traveling waves~\cite{Shen,epjp2020}.
This, in turn, renders particularly relevant and timely the systematic
revisiting of the analysis of shock waves in a prototypical integrable
discrete model, such as the Toda lattice.

The Toda lattice was discovered almost sixty years ago \cite{toda1967}, and has been studied extensively since then
(e.g., see \cite{blochkodama,deiftkamvissiskriecherbauerzhou,Teschl,toda1970,VDO} and references therein).
Much is known about it, including its direct and inverse spectral theory, its multi-soliton and finite-genus solutions, Whitham modulation equations, 
and solutions of various initial value problems. 
However, 
a complete%, self-contained and 
quantitative description of its periodic solutions and Whitham equations 
was never put together in a self-contained
form, to the best of our knowledge.
The purpose of this work is to remedy this situation as well as to present in detail the use of these tools to obtain a full characterization of the dispersive shocks of the system. Our hope and expectation
in subsequent work is that this will provide a helpful bridge towards
characterizing and understanding similar patterns in non-integrable
(and experimentally accessible) discrete systems. 
It may also provide a basis for similar exploration of other 
fully integrable systems, either semi-discrete
(i.e., continuous in time) or fully discrete in both space and time.

This presentation is structured as follows.
Interestingly, the one-dimensional Toda lattice appears in many different forms in the literature, 
and often with different and incompatible notations, hence in section~\ref{s:toda}
we begin by briefly reviewing them as well as presenting some basic properties of the system,
including its elliptic solutions and the so-called 
``Toda shock problem''.
In section~\ref{s:elliptic_whitham} we summarize the Toda-Whitham modulation equations
and provide explicit expressions of the characteristic speeds in the genus-one
case. In section~\ref{s:todadsw} we combine the theory of Secs.~\ref{s:toda}
and \ref{s:elliptic_whitham} to provide
a quantitative characterization of the dispersive shock waves of the Toda lattice.
Finally, in Sec.~\ref{s:simulations}, we provide detailed comparisons of
direct numerical simulations with the theory.

%%%%%%%%%%%%%%%%%%%%%%%%%%%%%%%%%%%%%%%%%%%%%%%%%%%%%%%%%%%%%%%%%%%%%
\section{The Toda lattice, its solutions and the Toda shock problem}
\label{s:toda}

\subsection{The Toda lattice}

The most direct way to write the Toda lattice is 
as the equation of motion for a chain of coupled 
nonlinear oscillators \cite{Toda}:
\vspace*{-0.4ex}
\[
M \ddot y_n = \e^{y_{n-1}-y_n} - \e^{y_n-y_{n+1}}\,,
\label{e:Toda_d2ydt2}
\]
that is,
\vspace*{-0.4ex}
\[
M \ddot y_n = - \phi'(y_n-y_{n-1}) + \phi'(y_n-y_{n+1})\,,
 \]
where a dot denotes differentiation with respect to time, $M$ is the mass, $y_n$ is the displacement
from the rest position, and where the potential function is 
\vspace*{-0.4ex}
\[
\phi(r) = 
\frac 1k\,(\e^{-kr} - 1\ ) + r
%\,\phi_o\,,
\]
with the mutual displacement between particles (i.e., the strain) defined as
\vspace*{-0.4ex}
\[
r_n = y_n - y_{n-1}\,. 
\label{e:y2r}
\]
Hereafter, boundary conditions are neglected, so we implicitly assume that the system holds
for all $n\in\Integer$, i.e., $N_2 = - N_1 = \infty$. 
Then \eqref{e:y2r} implies $y_n = y_{N_1} + \sum\nolimits_{k = N_1}^n r_k$.
Also, hereafter we set the constants
%$\phi_o$, 
$k$ and $M$ to 1 for simplicity, 
since the general system \eqref{e:Toda_d2ydt2} can be mapped into the canonical one by a suitable scaling transformation.

It is convenient to use the strains $r_n$ as the generalized coordinates for the system.  
%The generalized momenta are $s_n = \dot S_n$.
%
One can write~\eqref{e:Toda_d2ydt2} in Lagrangian or Hamiltonian form by introducing the Hamiltonian 
$H = K + U$,
where the kinetic and potential energies are, respectively,
\vspace*{-0.6ex}
\[
K = \half \sum_{n=N_1}^{N_2} \dot y_n^2\,,\quad
U = \sum_{n=N_1}^{N_2} \phi(r_n)\,.
\]
Then the generalized momenta [i.e., the canonically conjugate variables to the $r_n$] 
are $s_n = \partial L/\partial\dot r_n$, where $L(r,\dot r) = K - U$ is the Lagrangian of the system.
Moreover,
$H(r,s) = K + U = \sum\nolimits_{n=N_1}^{N_2}[\half (s_{n+1} - s_n)^2 + \phi(r_n) ]$, 
since \eqref{e:y2r} implies $\dot y_n = s_n - s_{n-1}$,
and therefore Hamilton's equations of motion for~\eqref{e:Toda_d2ydt2} are 
\vspace*{-0.6ex}
\bse
\label{e:Toda_Hamiltonian}
\begin{gather}
\dot r_n = \partialderiv H{s_n} = - (s_{n+1} - 2s_n + s_{n-1})\,,\qquad
\label{e:Toda_H1}
\\
\dot s_n = - \partialderiv H{r_n} = - \partialderiv{\phi(r)}{r}\bigg|_{r=r_n}\,. 
\label{e:Toda_H2}
\end{gather}
\ese
Solving~\eqref{e:Toda_H2} for $r_n$ as a function of $\dot s_n$
yields $r_n = - R(\dot s_n)$, 
with $R(\dot s) = \log(1 + \dot s)$.
Then~\eqref{e:Toda_H1} becomes
\[
\deriv{[R(\dot s_n)]}t = - (s_{n+1} - 2s_n + s_{n-1})\,,
\label{e:Toda_sdot}
\]
which is the dual form of the system~\eqref{e:Toda_d2ydt2}.
Moreover, by letting 
$w_n = \dot s_n$, 
so that 
$R(\dot s_n) = - \log(1+ w_n)$,
and differentiating~\eqref{e:Toda_sdot},
one also gets the form of the Toda lattice used in \cite{jpa2010} and many other works, namely
\vspace*{-0.6ex}
\[
\deriv[2]{ }t \log(1 + w_n) = w_{n+1} - 2w_n + w_{n-1}\,.
\label{e:Toda_wn}
\]

%\footnote{Note that $w_n \equiv f_n$ in \cite{Toda} and $w_n = V_n$ in \cite{jpa2010}.}
%\[
%y_n = S_{n-1} - S_n\,,\qquad 
%\]
%In particular, 
%\[
%w_n = \phi'(r_n)\,,\qquad 
%\]

\subsection{Flaschka variables and Lax pair}
Yet another formulation of the Toda lattice that is quite common, especially in works related to integrability, 
uses the so-called Flaschka variables, namely
\be
\~a_n = \half \dot y_n\,,\qquad 
\~b_n = \half \e^{(y_n-y_{n+1})/2}\,.
\label{e:abdefO}
\ee
(Note however that many different conventions and scalings exist for these variables. See also below.) 
%including \cite{blochkodama,flaschka2,Toda}.
%In particular, the names of the variables $a_n\&b_n$ in \cite{flaschka2,Toda} are exactly the opposite of those used here.
It is straightforward to verify that, in these variables,
the equations of motion~\eqref{e:Toda_d2ydt2} are
\vspace*{-0.6ex}
\[
\deriv{\~a_n}t = 2(\~b_{n-1}^2 - \~b_n^2)\,,
\qquad
\deriv{\~b_n}t = \~b_n\,(\~a_n - \~a_{n+1})\,.
\label{e:adotbdotO}
\]
(The strain $r_n$ is recovered from \eqref{e:abdefO} simply as $r_n = - 2\log(2\~b_{n-1})$.
The variables in Eqs.~\eqref{e:abdefO} and~\eqref{e:adotbdotO} are exactly the same as in~\cite{Toda} except for the switch $a_n=\~b_n$
$b_n= \~a_n$.)
The reason why the Flaschka variables are useful is that, as is easily verified by direct calculation, 
\eqref{e:adotbdotO} can be written in Lax form as
\vspace*{-0.6ex}
\[
\deriv{L}t = [B,L]\,, 
\label{e:TodaLaxform}
\]
where $[A,B] = AB-BA$ is the matrix commutator, and
$L$ and $B$ are the doubly infinite tridiagonal matrices
\bse
\begin{gather}
L = \begin{pmatrix} \ddots & \ddots \\[-1ex]
  \ddots & \!\!\~a_{n-1}\!\! & \~b_n \\
  & \~b_n & \~a_n & \!\!\~b_{n+1}\!\! \\[-1ex]
  & & \!\!\~b_{n+1}\!\! & \!\!\~a_{n+1}\!\! & \ddots \\[-1ex] 
  & &  & \ddots & \ddots
\end{pmatrix},
\qquad
B = \begin{pmatrix} \ddots & \ddots \\[-1ex]
 \ddots & 0 & \~b_n  \\
 & - \~b_n & 0 & \!\!\~b_{n+1}\!\! \\[-1ex]
 & & \!\! - \~b_{n+1}\!\! & 0 & \ddots \\[-1ex]
  & & & \ddots & \ddots
\end{pmatrix}.
\end{gather}
\ese
Note 
$L$ and $B$ can be thought of as operators in $l^2(\mathbb Z)$.
Specifically, 
\[
L = \~a_n I + \~b_{n+1}\,\e^\partial + \~b_n\,\e^{-\partial}\,,
\qquad
B = \~b_{n+1}\e^\partial - \~b_{n}\e^{-\partial}\,,
\]
where $I$ and $\e^\partial$ are respectively the identity and the shift operator, 
the latter acting as
$\e^\partial f_n = f_{n+1}$.
Of these operators, $L$
(which encodes the state of the system at time~$t$) is self-adjoint, while $B$ is skew-symmetric.  
The inverse scattering transform for the Toda lattice is based on the spectral theory of $L$,
i.e., the study of the eigenvalue problem $L v = \l v$ (e.g., see \cite{AblowitzPrinariTrubatch}).

For the purposes of the present work, 
it will be most convenient to use a different choice for the Flaschka variables
and to replace~\eqref{e:abdefO} by defining $a_n\&b_n$, as in \cite{blochkodama},
as
\be
a_n = - \dot y_n\,,\qquad 
b_n = \e^{y_{n-1} - y_n}\,,
\label{e:abdef}
\ee
which then replaces~\eqref{e:adotbdotO} with the ODEs
\be
\deriv{a_n}t = b_{n+1} - b_n\,,\qquad
\deriv{b_n}t = b_n(a_n-a_{n-1}).
\label{e:adotbdot}
\ee
(In the notation of \cite{blochkodama}, $a_n(t) = u_0(n,t)$ and $b_n(t) = u_1(n,t)$. 
The map between the two sets of dependent variables is obviously
$a_n, = - 2\~a_n$ and $b_n = 4 \~b_{n-1}^2$.)

\subsection{Soliton solutions}

The Toda lattice admits constant solutions in the form $a_n(t) = a$ and $b_n(t) = b$, with $a$ and $b$ 
arbitrary constants.
(One can verify that indeed in this case \eqref{e:TodaLaxform} yields $dL/dt = 0$.)
The Toda lattice also admits $N$-soliton solutions for all $N\in\Natural$.
For example, these solutions can be written in the form (e.g., see \cite{jpa2010})
\vspace*{-0.6ex}
\[
w_n(t) = \deriv[2]{ }t \log \tau(n,t)\,,
\]
where the $\tau$ function is given by
\vspace*{-0.6ex}
\begin{gather}
\tau(n,t) = \det\begin{pmatrix} 
   f_1(n,t) & \cdots & f_1(n+N-1,t) \\
   \vdots & \ddots & \vdots \\
   f_N(n,t) & \cdots & f_N(n+N-1,t)   
\end{pmatrix}\,,
\\
f_j(n,t) = \e^{\theta_j(n,t)} + (-1)^{j+1} \e^{- \theta_j(n,t)}\,,
\qquad
\theta_j(n,t) = \kappa_j n - \nu_j \sinh \kappa_j\,t + \theta_{j,0}\,,
\end{gather}
where $\nu_j = \pm 1$ and $\kappa_1<\cdots < \kappa_N$ are arbitrary.
In the simplest case $N=1$, one obtains the well-known solitary wave (i.e., one-soliton) solution of the Toda lattice:
\[
w_n(t) = \sinh^2\kappa\,\sech^2[\kappa n - \nu t \sinh\kappa]\,.
\]
(Note that in \cite{jpa2010} the signs $\nu_j$'s were all taken to be $-1$, and the variable $\kappa_j$ was called $\log p_j$.)

\subsection{Periodic traveling wave solutions of the Toda lattice}
\label{s:elliptic}

A partial family of periodic traveling wave elliptic solutions of the Toda lattice was first presented by Toda himself in \cite{toda1970}.
However, that is not the most general family of periodic solutions, as it only contains two free parameters, 
and all those solutions had zero average.
A full, four-parameter family of elliptic solutions of the Toda lattice was derived by Teschl \cite{Teschl}
as a special case of the finite-genus formalism.
%The equations of motion [(12.4) in \cite{Teschl}] are the same as \eqref{e:Toda_d2ydt2}.
Note that Teschl writes the Toda lattice in Flaschka variables as $\dot a_{n,\T} = a_{n,\T}\,(b_{n+1,\T}-b_{n,\T})$ and $\dot b_{n,\T} = (a_{n,\T}^2 - a_{n-1,\T}^2)$,
by introducing the variables $a_{n,\T} = \half \e^{(y_n-y_{n+1})/2}$ and $b_{n,\T} = - \half \dot y_n$,
where we employed the subscript ``T'' to denote the Flaschka variables in the notation of \cite{Teschl} to avoid confusion with the notation used in the present work.)
Therefore, the map between the two sets of dependent variables is $a_{n,\mathrm{T}} = \~b_n$ and $b_{n,\mathrm{T}} = - \~a_n$, 
implying $a_n =2 b_{n,\mathrm{T}}$ and $b_n = 4a_{n-1,\mathrm{T}}^2$.

Explicitly, following \cite{Teschl}, the general four-parameter family of periodic solution of the Toda lattice is given by
\newpage 
%Teschl obtains the following solutions in terms of $E_1,\dots,E_4$, 
\bse
\begin{gather}
b_{n+1}(t) = \half(E_1^2 + E_2^2 + E_3^2 + E_4^2) + 2\^R_n(t) - \mu_n^2(t) - \txtfrac1{4} a_n^2(t)\,,
\label{e:aelliptic}
\\
a_n(t) = E_1 + E_2 + E_3 + E_4 - 2\mu_n(t)\,,
\label{e:belliptic}
\end{gather}
\ese
where
\vspace*{-1ex}
\bse
\begin{gather}
\mu_n(t) = E_2 \frac{\displaystyle 1 - (E_1/E_2)\,B\,\sn^2(Z_n(t),m)}{1 - B\,\sn^2(Z_n(t),m)}\,,
\\
Z_n(t) = 2n F(\Delta,m) + \omega t\ + Z_0, \label{e:Z}
\\
m = \frac{(E_3-E_2)(E_4-E_1)}{(E_4-E_2)(E_3-E_1)}\,,
\label{e:mdef}
\\
\hat R_n(t) = - \sigma_n(t) \,\sqrt{P(\mu_n(t))}\,,\qquad
P(z) = (z-E_1)(z-E_2)(z-E_3)(z-E_4)\,,
\\
\omega = \sqrt{(E_4-E_2)(E_3-E_1)}\,,\qquad
\Delta = \sqrt{\frac{E_4-E_2}{E_4-E_1}}\,,\qquad
B = \frac{E_3-E_2}{E_3-E_1}\,.
\end{gather}
\ese
In the above equations,
$E_1,\dots,E_4$ are the branch points of the genus-one Riemann surface associated with the above solution,
with $E_1<E_2<E_3<E_4$;
$\sn(z,m)$ is the Jacobi elliptic sine
(see section~\ref{s:simulations} for further details), 
$F(z,m)$ is the inverse of $\sn(z,m)$, 
$Z_0$ is an arbitrary translational parameter (phase),
 % The slow variation of the phase is not determined by leading-order Whitham theory. 
$\mu_n(t)$ is the Dirichlet eigenvalue of the scattering problem for the Toda lattice,
and $\sigma_n(t) = \pm1$ is the sign associated with $\mu_n(t)$, and determines whether $\mu_n(t)$ is increasing or decreasing as a function of $n$ and $t$.
Note also that: 
(i) $\mu_n(t) \in [E_2,E_3]$ (the spectral gap of the scattering problem)
for all $(n,t)\in\Real^2$;
(ii) $\mu_n(t)$ achieves its minimum and maximum values (i.e., $E_2$ and $E_3$, respectively) at $Z_n(t) = 0$ and at $Z_n(t) = K(m)$,
respectively,
where $K(m)$ is the complete elliptic integral of the first kind;
(iii) the mean value of $\mu_n(t)$ over one period is
\[
\mu_\mathrm{mean} =  E_1 + (E_2-E_1)\,\frac{\Pi(B,m)}{K(m)}\,,
\]
where $\Pi(\,\cdot\,)$ is the complete elliptic integral of the third kind.
Accordingly, the minimum and maximum values of $a_n(t)$ are, respectively,
\[
a_\mathrm{min} = \half\,(E_1 + E_2 - E_3 + E_4)\,,
\qquad
a_\mathrm{max} = \half\,(E_1 - E_2 + E_3 + E_4)\,, \label{e:bextreme}
\]  
and are reached respectively at $Z_n(t) = K(m)$ and $Z_n(t)=0$, and its mean value is
\[
a_\mathrm{mean} = \half\,(E_2+E_3+E_4-E_1) - (E_2-E_1)\,\frac{\Pi(B,m)}{K(m)}\,.
\label{e:mean}
\] 

The harmonic limit ($m\to0^+$) of the above solution is obtained as $E_3\to E_2^+$.
In this case,
similarly to what happens for the KdV equation,
the solution reduces to small-amplitude harmonic oscillations.
(Recall $\sn(z,0) = \sin(z)$,
$F(\Delta,0) = \arcsin\Delta$,
$K(m=0) = \pi/2$
and 
%$\Pi(z,0) = \pi/(2\sqrt{1-z})$}:
$\mu_n(t) \to E_2 + (E_3-E_2)\,\sin^2(Z_n(t)) + O(E_3-E_2)^2$.)
Conversely, the soliton limit ($m\to1^-$) of the above solution is obtained as $E_3\to E_4^-$.
In this limit, the above expression yields the one-soliton solution of the Toda lattice.
(Recall $\sn(z,1) = \tanh z$, 
$F(\Delta,1) = \log[(1+\Delta)/\sqrt{1-\Delta^2}]$ 
and 
%$K(m)\to\infty$
%$K(m) = \half\log[16/(1-m)] - \frac18(1-m)\,[\log(1-m) + 2 - 4\log2] + O(1-m)^2$
$K(m) = \half\log[16/(1-m)] - \frac18(1-m)\log(1-m) + O(1-m)$
as $m\to1$.)

Additionally, since the Toda lattice is an integrable system, it also admits finite-genus solutions of arbitrary genus,
which were written down in \cite{Teschl}.  
However, those are not needed here, so for brevity we will not discuss them.

\subsection{The Toda shock problem and the Toda rarefaction problem}
The Toda shock problem,  first considered by Holian, Flaschka and McLaughlin in~\cite{PRB1981v24p2595}, 
and then studied by Venakides, Deift and Oba using IST methods in \cite{VDO},
is the IVP for~\eqref{e:Toda_d2ydt2} with ICs
\vspace*{-0.6ex}
\[
y_n(0) = 0\,,\qquad 
\dot y_n(0) = -2c\,\sign(n)\,,
\label{e:TodashockIC}
\]
with $c>0$ and $\sign(0)=0$.
(As noted in \cite{VDO}, more general ICs of the form $y_n(0) = d n + e$ can be reduced to the above ones 
by a trivial transformation.)
The Toda rarefaction problem is the above IVP but with $c<0$, and was studied by Deift, Kamvissis and Kriecherbauer
in \cite{deiftkamvissiskriecherbauerzhou}.
Both the shock problem and the rarefaction problem were then studied in \cite{blochkodama} 
within the framework of Whitham modulation theory.
In the following sections we will focus our attention on the Toda shock problem, since this is the case that leads to the formation of dispersive shocks.
%
\iffalse
The dynamics of solutions for the Toda shock problem differs depending on whether $0<c<1$ or $c>1$.
Specifically:
(i)~if $0<c<1$, $y_n(t)$ tends to a finite limit as $t\to\infty$;
(ii)~if $c>1$, $y_n(t)$ limits to an oscillatory motion as $t\to\infty$.
\fi

%%%%%%%%%%%%%%%%%%%%%%%%%%%%%%%%%%%%%%%%%%%%%%%%%%%%%%%%%%%%%%%%%%%%%
\section{Whitham modulation equations for the Toda lattice}
\label{s:elliptic_whitham}

\subsection{General formulation of the Whitham equations for the Toda lattice}

Since the Toda lattice is an integrable system, it possesses an infinite number of conservation laws,
and as a result one can write down Whitham equations of arbitrary genus
which govern the modulations of the corresponding finite-genus solutions of the system. 
These Whitham equations were written down in an elegant form by Bloch and Kodama in \cite{blochkodama} 
by making use of the integrability structure of the Toda lattice.

Specifically, and without repeating the derivation, 
it was shown in \cite{blochkodama}
that the genus-$g$ Whitham equations for the Toda lattice 
in diagonal (i.e., Riemann invariant) form are given by
\[
\partialderiv{\l_j} t - s_j(\fbf\l)\,\partialderiv{\l_j}x = 0\,,\qquad j = 1,\dots,2g+2\,, 
\label{e:TodaWhitham}
\]
with $\fbf\l = (\l_1,\dots,\l_{2g+2})$,
where the Riemann invariants $\l_1,\dots,\l_{2g+2}$ are the branch points of the spectral curve associated with 
the genus-$g$ solutions of the Toda lattice.
The characteristic speeds $s_1,\dots,s_{2g+2}$ in~\eqref{e:TodaWhitham} are given by $s_j(\fbf\l) = S(\fbf\l,\l_j)$, 
where the unique function $S(\fbf\l,z)$ is %expressed as
%[suppressing the dependence on $\fbf\l$]
\[
S(\fbf\l,z) =  \frac{ z^{g+1}  + \gamma_o z^g + \gamma_1 z^{g-1} + \cdots + \gamma_g }
  {2\prod\nolimits_{j=1}^g (z - \alpha_j)}\,,
\label{e:characteristicspeeds}
\]
with $\gamma_o  = - \half\sigma_1$ and $\sigma_1 = \l_1 + \cdots + \l_{2g+2}$, and 
where the coefficients $\alpha_1,\dots,\alpha_g$ and $\gamma_1,\dots,\gamma_g$ 
(which depend on all the quantities $\l_1,\dots,\l_{2g+2}$) 
are given by the solution of the $2g\times2g$ system of equations%
\bse
\label{e:Whithamcoefficientsystem}
\begin{gather}
I_g^k + P_1 I_g^{k-1} + \cdots + P_g I_g^0 = 0\,,\qquad k = 1,\dots,g\,,
\label{e:Whithamcoefficientsystem1}
\\
I_g^{k+1} + \gamma_o I_g^k + \gamma_1 I_g^{k-1} + \cdots \gamma_g I_g^0 = 0\,,\qquad k = 1,\dots,g\,,
\label{e:Whithamcoefficientsystem2}
\\
\noalign{\noindent with}
P_1 = - \sum_{k=1}^g\alpha_k~,\qquad
P_2 = \sum_{k_1=1}^g\sum_{k_2=k_1+1}^g\alpha_{k_1}\alpha_{k_2}~\quad\cdots~\quad
P_g = (-1)^g\,\prod_{k=1}^g\alpha_k\,,
\\
I_k^j = \int_{\l_{2k}}^{\l_{2k+1}} \frac{\l^j\,\d\l}{\sqrt{R_g(\l)}}\,,
\\
R_g(\l) = \prod_{k=1}^{2g+2}(\l - \l_k)\,.
\end{gather}
\ese
The system~\eqref{e:Whithamcoefficientsystem} arises from the normalization conditions of the Abelian differentials associated with the genus-$g$ solutions
of the Toda lattice.

It was shown in \cite{blochkodama} that the characteristic speeds satisfy a ``sorting property'', namely:
(i)~ for all $j$ and $k$ such that $\l_j<\l_k$, one has $s(\l_j) < s(\l_k)$;
(ii)~ $\partial[s(\l_j)]/\partial\l_j>0$ for all $j=1,\dots,2g+2$.
This property is crucial in order to ascertain the existence or non-existence of global solutions to the IVP for the Whitham equations
(e.g., see \cite{JNLS16p435,blochkodama,SIAP59p2162}).
Specifically, in \cite{blochkodama} Bloch and Kodama proved that 
if the ICs for the system \eqref{e:TodaWhitham} are non-increasing, the IVP has a global solution.

In order to apply the above Whitham equations to study modulations of the periodic solutions discussed in the previous section, 
one obviously needs a map between the Riemann invariants in Whitham theory of~\cite{blochkodama} with $g=1$
and the branch points in the elliptic solution of~\cite{Teschl}.
This map can be obtained by recalling that, 
on one hand, 
$a_n = -2 \~a_n = 2 b_{n,\mathrm{T}}$ and 
$b_n = 4\~b_{n-1}^2 = 4a_{n-1,\mathrm{T}}^2$,
and, on the other hand,
the variable $a_n$ in section~\ref{s:elliptic} is directly proportional to the $E_j$'s, while $a_n$ in \cite{blochkodama} is proportional to the $\l_j$'s
(also see section~\ref{s:genus0whitham} below).
Therefore one simply has $\l_j = 2E_j$ for $j=1,\dots,4$.

\subsection{Genus-zero Whitham equations}
\label{s:genus0whitham}
The genus-zero Whitham system for the Toda lattice 
is the special case $g=0$ of the general system~\eqref{e:TodaWhitham} above,
and describes the modulations of the ``constant'' solutions of the Toda lattice.
In the normalization of \cite{blochkodama},
one has (cf.~(4.2) in \cite{blochkodama})
\[
a_n = \half(\l_1 + \l_2),
\qquad
b_n = \txtfrac1{16}(\l_1-\l_2)^2,
\]
and the characteristic speeds are 
\[
s_1 = - \sqrt{b_n} = \txtfrac14 (\l_1-\l_2)\,,
\qquad
s_2 = \sqrt{b_n} = - \txtfrac14 (\l_1-\l_2)\,.
\]

\subsection{Genus-one Whitham equations}
The genus-one Whitham equations are the equations that govern the modulations of the periodic solutions of the Toda lattice,
and, as we show below, such equations are the most useful ones for characterizing the dispersive shock waves in the problem of interest.
In the case $g=1$, the expression~\eqref{e:characteristicspeeds} for the characteristic speeds simplifies to
\be
S(\fbf\l,z) = \frac{z^2 + \gamma_o z + \gamma_1}{2(z - \alpha_1)}\,, 
\label{e:g1speeds}
\ee
where the constants $\alpha_1$ and $\gamma_1$ are determined by the solution of the linear system 
\eqref{e:Whithamcoefficientsystem},
which for $g=1$ reduces to 
\bse
\begin{gather}
I_1^1 - \alpha_1 I_1^0 = 0\,,
\\ 
I_1^2 + \gamma_o I_1^1 + \gamma_1 I_1^0 = 0\,,
\end{gather}
\ese
which immediately implies
\be
\alpha_1 = I_1^1/I_1^0\,,\qquad
\gamma_1 = - (I_1^2 + \gamma_o I_1^1)/I_1^0\,. 
\label{e:g1coeffs}
\ee
The coefficients in the above system are given by the following elliptic integrals:
%
%\bse
\begin{gather}
I_1^j = \int_{\l_2}^{\l_3} \frac {\l^j}{\sqrt{R_1(\l)}}\,\d\l\,,
\qquad
R_1(\l) = (\l - \l_1)(\l - \l_2)(\l - \l_3)(\l - \l_4)\,.
\label{e:I1j}
\end{gather}
%\ese

\subsection{Explicit expressions for the characteristic speeds for the genus-one Toda-Whitham equations}

Notably, all the integrals in Eq.~\eqref{e:I1j} can be evaluated exactly, using in particular formulae 
250.00, 250.01, 254.00, 254.10 or 255.00 (see also 340.04) in \cite{ByrdFriedman}.
%(Note that \cite{ByrdFriedman} uses $a>b>c>d$ for the roots, so $a = \l_4,~ b = \l_3,~ c = \l_2,~ d = \l_1$.)
Explicitly:
\bse
\begin{gather}
I_1^0 = G K(m)\,,
\\
I_1^1 = G \,[ \l_1 K(m) + (\l_2 - \l_1)\,\Pi_m ]\,,
\\
I_1^2 = G \,\bigg[ 
  \l_1^2 K(m) + 2 \l_1\,(\l_2 - \l_1)\,\Pi_m
\nonumber\kern26em
\\
\kern6em
  - \frac{(\l_2 - \l_1)^2}{2(1-\alpha^2)(m-\alpha^2)}
      \big( \alpha^2 E(m) + (m - \alpha^2)\,K(m) + (2 \alpha^2 m + 2 \alpha^2 - \alpha^4 - 3m)\,\Pi_m\big)
    \bigg]\,,
\end{gather}
\ese
where
$E(m)$ is the complete elliptic integral of the second kind,
and
\begin{gather}
m = \frac{(\l_3-\l_2)(\l_4-\l_1)}{(\l_4-\l_2)(\l_3-\l_1)}\,,\qquad
\alpha^2 = \frac{\l_3-\l_2}{\l_3-\l_1}<m\,,\qquad
G = \frac2{\sqrt{(\l_4-\l_2)(\l_3-\l_1)}}\,.
%\beta = \frac{\l_1}{\l_2}\,,
\label{e:malphag}
%\\
%V_2 = - \frac1{2(1-\alpha)(m-\alpha)}\,\big[ \alpha E(m) + (m - \alpha)\,K(m) + 
%  (2 \alpha m + 2 \alpha - \alpha^4 - 3m)\,\Pi_m(\alpha) \big]\,.
\end{gather}
Here, for brevity $\Pi_m = \Pi(\alpha^2,m)$, where as before $\Pi(\,\cdot\,)$ is the complete elliptic integral of the third kind.
%(Note that $\alpha = \alpha^2_\mathrm{BF}$, $\beta = \alpha_1^2/\alpha^2 = \l_1/\l_2$ in \cite{ByrdFriedman}.)
%
Substituting $I_1^0,I_1^1\&I_1^2$ in~\eqref{e:g1coeffs} and the resulting expressions for $\alpha_1$ and $\gamma_1$ 
in turn into~\eqref{e:g1speeds}
yields, after some tedious but straightforward algebra, explicit expressions for the characteristic speeds of the genus-one Toda-Whitham equations, 
which are novel to the best of our knowledge.
Specifically,
\be
S(\fbf\l,z) = \big[ (2z^2 -\Sigma z + \l_2\l_3 + \l_1 \l_4 )\,K(m) + (\l_3-\l_1)(\l_4-\l_1) \big] 
  \big/\big[ 4((z - \l_1)K(m) - (\l_2-\l_1)\Pi_m) \big]\,,
\label{e:genus1speeds} 
\ee
where $\Sigma = \l_1 + \l_2 + \l_3 + \l_4$
and $m$ and $\alpha$ are given by \eqref{e:malphag}.
Then, substituting $\l = \l_j$ in \eqref{e:genus1speeds} for $j=1,\dots,4$, one then finally obtains, explicitly:
%\vspace*{-0.6ex}
\bse
\label{e:s1s4}
\begin{gather}
s_1(\fbf\l) =  - \frac{ (\l_3-\l_1) ( (\l_4-\l_2)E(m) + (\l_2-\l_1)K(m)) }{4 (\l_2-\l_1) \Pi_m}\,,
\\
s_2(\fbf\l) = \frac{ (\l_4-\l_2) ( (\l_3-\l_1)E(m)-(\l_2-\l_1)K(m)) }{4 (\l_2-\l_1) (K(m)-\Pi_m)}\,,
\\
s_3(\fbf\l) = - \frac{ (\l_3-\l_1) ( (\l_4-\l_3)K(m) - (\l_4-\l_2)E(m) ) }{4 ( (\l_3-\l_1)K(m) - (\l_2-\l_1)\Pi_m)}\,,
\\
s_4(\fbf\l) = \frac{ (\l_4-\l_2) ( (\l_4-\l_3)K(m) - (\l_3-\l_1)E(m) ) }{4 ((\l_4-\l_1)K(m) - (\l_2-\l_1)\Pi_m)}\,.
\end{gather}
\ese
These explicit and simplified expressions for the characteristic speeds will
be crucial in applying the Whitham theory for the practical
description of DSWs, detailed in below in Sec.~\ref{s:todadsw}
and \ref{s:simulations}.
Note how the characteristic speeds \eqref{e:s1s4} contain the complete elliptic integral of the third kind, 
unlike those for the genus-one Whitham equations for the Korteweg-deVries and nonlinear Schr\"odinger equations~\cite{Mark2016}.

\subsection{Harmonic limit and soliton limit}

Like other genus-one Whitham systems, the above Toda-Whitham modulation system of equations admits 
distinguished limits as $m\to0$ and $m\to1$.
Specifically, the harmonic limit, $m\to0$, is obtained when $\lambda_3\to\lambda_2^+$, 
in which case the elliptic solution of the Toda lattice reduces to a vanishing-amplitude harmonic wave.
Recall the Taylor series expansion \cite{NIST2010}
\be
\Pi_m = \frac\pi2 \sum_{n=0}^\infty \frac{\big(\half\big)_n}{n!}
  \sum_{s=0}^n\frac{\big(\half\big)_s}{s!} m^s \alpha^{2n-2s}\,.
\label{e:Piseries}
\ee
Using~\eqref{e:Piseries}, one can see that $s_3(\fbf\l) = s_2(\fbf\l)$ as $m\to0$, 
and~\eqref{e:TodaWhitham} limits to a reduced system of modulation equations for $\l_1$, $\l_2$ and $\l_4$,
with
\begin{gather}
s_1 = - \txtfrac14(\l_4-\l_1),\quad
s_2 = 0,\qquad
s_4 = \txtfrac14(\l_4-\l_1).
\end{gather}
The corresponding system describes the propagation of a small-amplitude harmonic wave riding on top of the 
constant solution of the Toda lattice.

Conversely, the soliton limit, $m\to1$, is obtained when $\l_3\to\l_4^-$,
in which case the elliptic solution of the Toda lattice reduces to its soliton solution.
Using 111.04 from~\cite{ByrdFriedman}, we obtain
\be
\Pi_m = \frac{K(m)}{1-\alpha^2} - \frac{\alpha}{2(1-\alpha^2)}\log\bigg(\frac{1+\alpha}{1-\alpha}\bigg) + o(1-m),\qquad m\to1^-\,.
\ee
Using this expression one can see that $s_3(\fbf\l) = s_4(\fbf\l)$ in the limit $m\to1$, 
and~\eqref{e:TodaWhitham} again limits to a reduced system of modulation equations for $\l_1,\l_2\&\l_4$,
except that now
\bse
\begin{gather}
s_1 = - \txtfrac14(\l_2-\l_1),\qquad 
s_2 = \txtfrac14(\l_2-\l_1),\qquad
s_4 = \frac{\sqrt{(\l_4-\l_1)(\l_4-\l_2)}}{2\displaystyle\log\bigg(\frac{\sqrt{\l_4-\l_1}+\sqrt{\l_4-\l_2}}{\sqrt{\l_4-\l_1}-\sqrt{\l_4-\l_2}}\bigg)}\,.
\end{gather}
\ese

The soliton limit of the genus-one Whitham equations can now be used to study 
the dynamics of a soliton propagating on a slowly varying background, e.g., as in 
\cite{Hoefer2018,Hoefer2018a,NLTY2021v34p3583}.
Similarly, the harmonic limit of the genus-one Whitham equations will be useful to study 
the dynamics of a small-amplitude harmonic wave propagating on a slowly varying background.

%%%%%%%%%%%%%%%%%%%%%%%%%%%%%%%%%%%%%%%%%%%%%%%%%%%%%%%%%%%%%%%%%%%%%
\section{Use of the elliptic solutions and Whitham equations to characterize the dispersive shocks}
\label{s:todadsw}

\subsection{Analysis of the Toda shock problem and rarefaction problem via the Whitham equations}
In~\cite{blochkodama}, 
Bloch and Kodama used Whitham theory to study the Toda shock problem and 
Toda rarefaction problem, which, expressed in terms of the Flaschka variables, are
\[
a_n(0) = 2c\,\sign(n),\qquad b_n(0) = 1\,.
\label{e:ICblochkodama}
\]
For the genus-zero system above, in the notation of \cite{blochkodama}, 
the corresponding ICs are then
\[
\l_1(x,0) = \begin{cases} -2(c+1),&x<0, \\ 2(c-1),&x>0,\end{cases}\qquad
\l_2(x,0) = \begin{cases} - 2(c-1), &x<0, \\ 2(c + 1), &x>0. \end{cases}
\label{e:ICblochkodama_g0}
\]
It was then shown in \cite{blochkodama} that if $c>0$, the genus-zero system
with the above ICs develops shocks,
whereas if $c<0$ the system does not develop shocks.
More precisely, in \cite{blochkodama} it was shown that:
(i)~If $c>1$, the ICs~\eqref{e:ICblochkodama} are regularized by genus-one data.
(ii)~If $0<c<1$, the ICs~\eqref{e:ICblochkodama} are regularized by genus-two data.
(iii)~If $c<0$, the ICs~\eqref{e:ICblochkodama} do not break, the genus-zero system has a global solution.
A concise description of the time evolution of the Riemann invariants arising in each case,
together with the evaluation of the characteristic speeds of the boundaries between regions of different genus,
is found in \cite{blochkodama}. 
A much more detailed analysis of each case,
including  descriptions of the full solution of the Whitham modulation system and of the corresponding solution of the Toda lattice,
as well as comparisons with numerical simulations,
are provided below.

\subsection{Practical guide for characterizing the Toda shock problem via Whitham modulation theory}

In this section we bring together the theory detailed above
in order to obtain concrete analytical predictions of various
characteristics of the DSWs of the Toda lattice. Such predictions
will be compared to numerical simulations of the full Toda lattice
problem in the subsequent section. 

Recall that if the ICs for a given system of Whitham modulation equations are non-increasing, the system 
admits a global solution.
As usual when using Whitham modulation theory to analyze Riemann problems, 
if the ICs for the genus-zero system are not non-increasing, 
one then tries to ``regularize them'' by writing them as a degenerate case of a higher-genus system
(e.g., see \cite{JNLS16p435,blochkodama,PHYSD236p44,SIAP59p2162}).
In the case of the Toda shock problem, the precise steps involved (and the corresponding results)
depend crucially on the value of the speed $c$, so one needs consider the following two cases seperately:
(i) $c>1$ and (ii) $0<c<1$.
(We will not discuss the case $c<0$ since in this case no DSWs are produced.)

For each of the two cases, 
to compare the predictions of Whitham modulation theory to direct numerical simulations of the Toda lattice,
%involves, 
it is convenient to proceed according to the following steps:
\vspace*{-1ex}
\begin{enumerate}
\item 
On the density plots of the solution as a function of $n$ and~$t$, draw straight lines representing the boundaries of the oscillation zones.
\item 
Given the fixed values 3 of the 4 Riemann invariants, evaluate the elliptic integrals as a function of the fourth one.
Then, at fixed $t$, use the value of the relevant integral to express $x$ explicitly as a function of~$m$.
\item 
In the temporal snapshots of the solution use the self-similar solutions of the Whitham equations and the corresponding max/min of the elliptic solutions to plot the envelope of the oscillations as a function of~$n$. % CC: I changed m to n here
\item 
Using the formulae for the elliptic solutions, plot the full oscillations as a function of~$n$.
\end{enumerate}
Next we discuss these four steps in more detail for each of the two cases. 

\subsection{Case 1: $c>1$} \label{sec:case1}

\begin{figure}[t!]
\centerline{
   \begin{tabular}{@{}p{0.4\linewidth}@{}p{0.4\linewidth}@{}}
     \rlap{\hspace*{5pt}\raisebox{\dimexpr\ht1-.1\baselineskip}{\bf (a)}}
 \includegraphics[height=5cm]{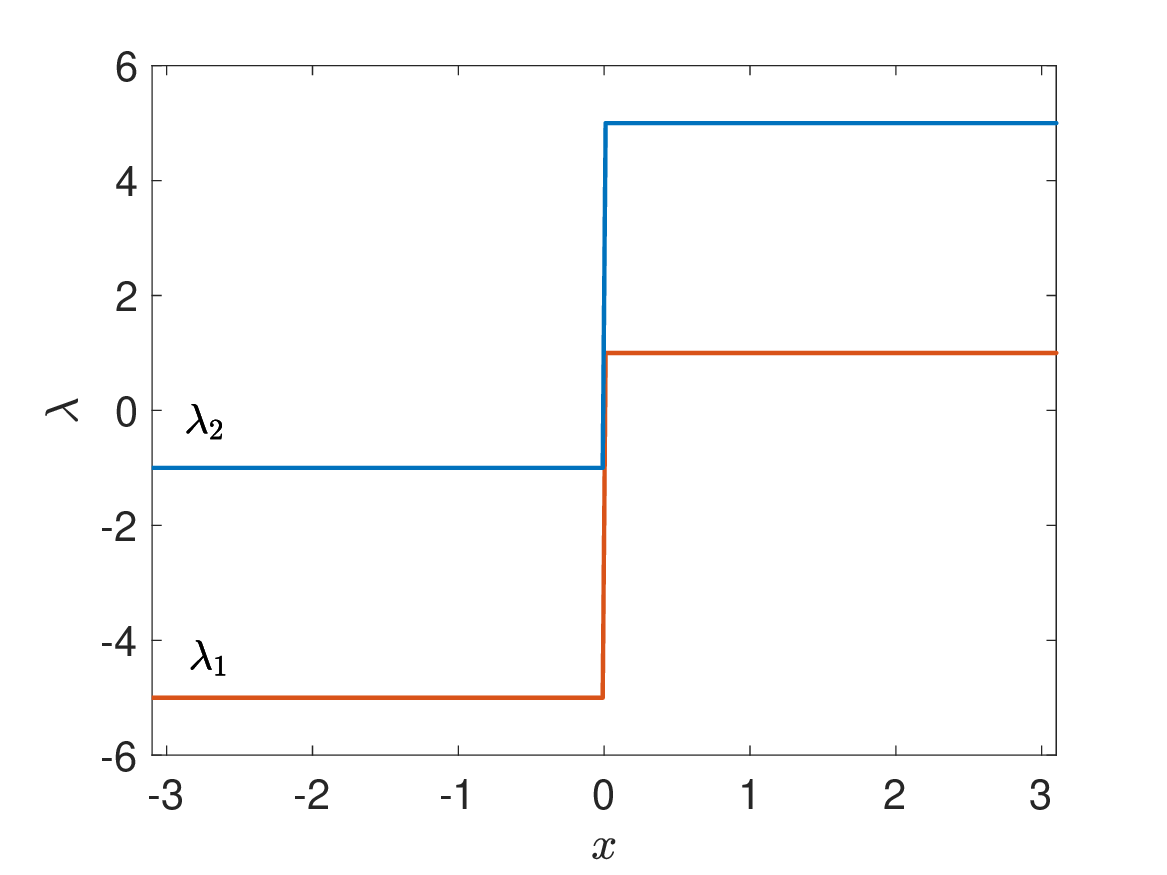} &
  \rlap{\hspace*{5pt}\raisebox{\dimexpr\ht1-.1\baselineskip}{\bf (b)}}
 \includegraphics[height=5cm]{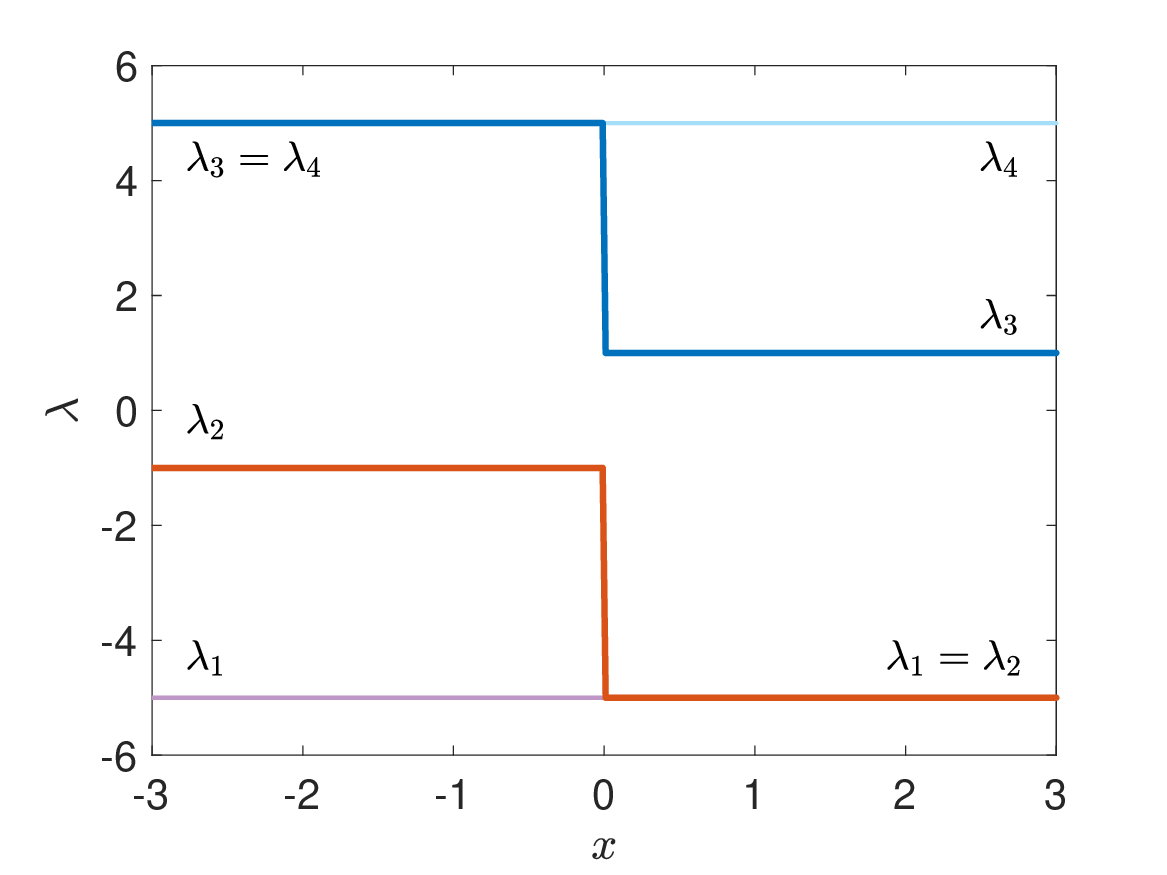} 
  \end{tabular}
  }
\centerline{
   \begin{tabular}{@{}p{0.4\linewidth}@{}p{0.4\linewidth}@{}}
     \rlap{\hspace*{5pt}\raisebox{\dimexpr\ht1-.1\baselineskip}{\bf (c)}}
 \includegraphics[height=5cm]{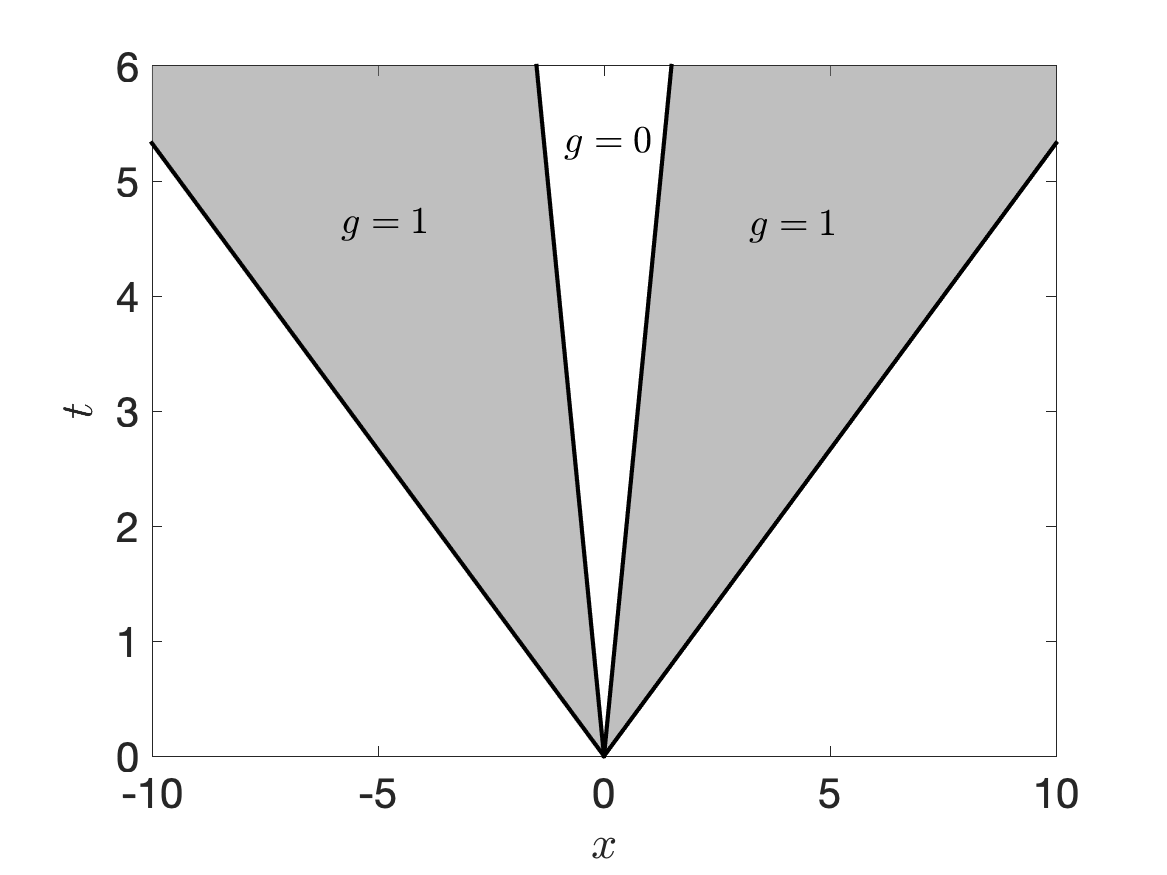} &
  \rlap{\hspace*{5pt}\raisebox{\dimexpr\ht1-.1\baselineskip}{\bf (d)}}
 \includegraphics[height=5cm]{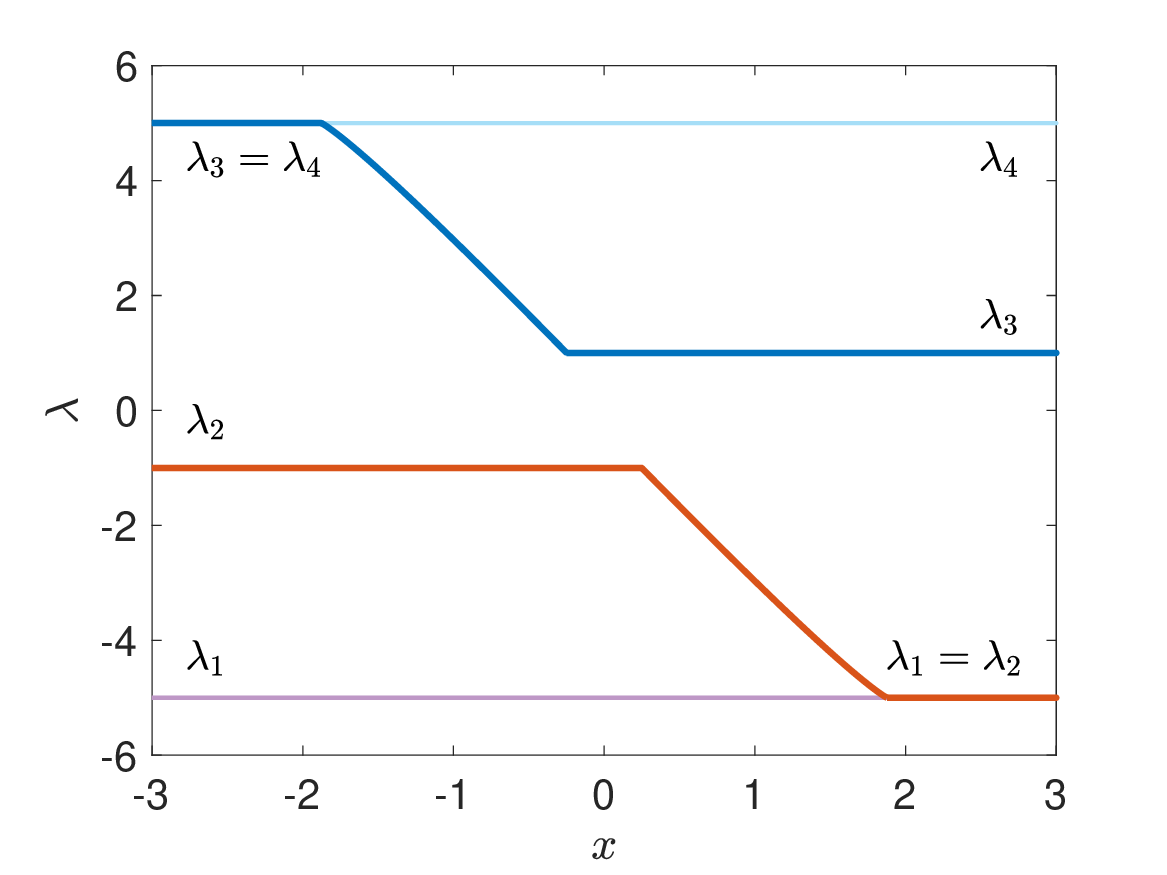} 
  \end{tabular}
  }
\caption{%
\textbf{(a)} ICs for the genus-zero Whitham system %(dark blue: $\l_2$; orange: $\l_1$) 
for case~1 ($c=3/2$).
\textbf{(b)} Regularization of the IC via the genus-one Whitham system.
\textbf{(c)} Bifurcation plot showing the genus-one regions (in light gray) and genus-zero regions (in white) in the $xt$-plane. The core of the DSWs will be located in the gray regions,
which will be counter-propogating. 
\textbf{(d)} The value of the Riemann invariants at $t=1$ (see text for details).
}
\label{f:1}
\end{figure}

\paragraph{Step~1.}
Recall that, 
when $c>1$, the IC for $a_n$ and $b_n$, given in~\eqref{e:ICblochkodama} are regularized by genus-one data.
To see why this is the case, note that the ICs for the genus-zero Whitham equations corresponding to those in \eqref{e:ICblochkodama} are given by~\eqref{e:ICblochkodama_g0}, and are shown in Fig.~\ref{f:1}(a) (cf.~Fig.1 in~\cite{blochkodama}).
Since these IC are not non-increasing, the time evolution of the genus-zero system gives rise to a shock.
The shock is regularized by embedding the IVP into a degenerate case of the genus-one Whitham equations, 
for which the corresponding IC are as follows:
\vspace*{-0.4ex}
\bse
\label{e:Case1_IC_g1}
\begin{gather}
\l_1(x,0) =  - \l_4(x,0) = -2(c+1),\quad \forall x\in\Real\,,
\\
\l_2(x,0) = \left\{\kern-0.4em\begin{array}{ll}-2(c-1),&x<0,\\ -2(c+1) = \l_1,&x>0,\end{array}\right.
\qquad
\l_3(x,0) = \left\{\kern-0.4em\begin{array}{ll}2(c+1) = \l_4,&x<0,\\ 2(c-1),&x>0,\end{array}\right.
\end{gather}
\ese
These IC, 
shown in Fig.~\ref{f:1}(b) (cf.~(4.6) and Fig.2 in~\cite{blochkodama}),
are non-increasing, and therefore the genus-one Whitham system admits a global solution.
In particular, %(see the cartoon in~Fig.3 in~\cite{blochkodama}):
\bse
\vspace*{-0.4ex}
\begin{gather}
\l_1(x,t) =  - \l_4(x,t) = -2(c+1),\quad \forall (x,t)\in\Real\times\Real^+\,,
\\
\l_2(x,t) = \left\{\kern-0.4em\begin{array}{ll}-2(c-1), &x < s_2^-t,\\ -2(c+1) = \l_1, &x > s_2^+t,\end{array}\right.
\qquad
\l_3(x,t) = \left\{\kern-0.4em\begin{array}{ll}2(c+1) = \l_4, &x<s_3^-t,\\ 2(c-1),&x>s_3^+t.\end{array}\right.
\label{e:solution_c>1b}
\end{gather}
\ese
The value of $\l_2$ in the transition region $s_2^-t < x < s_2^+t$ varies continuously from $\l_2 = \l_2^- = -2(c-1)$
to $\l_2 = \l_2^+ = \l_1 = -2(c+1)$.
Similarly, the value of $\l_3$ in the transition region $-s_3^-t < x < - s_3^+t$ varies continuously from $\l_3 = \l_3^- = 2(c+1)$ to $\l_3 = \l_3^+ = 2(c-1)$.
\footnote{The values of $\lambda_2,\lambda_3$ in the transition region are depicted qualitatively in~Fig.~3 of~\cite{blochkodama}. We give a quantitative description of these quantities below in Eq.~\eqref{e:lambda3case1} and Fig.~\ref{f:1}(d).}
Therefore, the time evolution gives rise to two DSWs, located in the expanding regions 
$-s_3^-t < x < - s_3^+t$ and
$s_2^-t < x < s_2^+t$, in which the values of all 4 Riemann invariants are distinct from each other, see Fig.~\ref{f:1}(c)
(and also Fig.~3 in \cite{blochkodama}). 
Bloch and Kodama compute the speeds of the boundaries between the regions of different genus in the various cases studied.
The values of $s_3^\pm$ and $s_2^\pm$ are 
%obtained by computing suitable limits of the integrals defined earlier. 
%\\
%Specifically B\&K write
(cf.~(4.15) and (4.16) in \cite{blochkodama}, but note that (4.16) in \cite{blochkodama} is off by a minus sign):
\vspace*{-0.4ex}
\bse
\begin{gather}
s_3^- = - s_2^+ = \frac{\sqrt{c(c+1)}}{\log(\sqrt{c}+\sqrt{c+1})} > 1\,, \label{e:s3-}
\qquad
s_3^+ = - s_2^- = 2(c-1)(1-\Gamma) > 0\,, 
\\
\Gamma = {\int_0^{\pi/2}\frac{\sin^4\phi}{\sqrt{(c\cos^2\phi+\sin^2\phi)(c\sin^2\phi+\cos^2\phi)}}\d\phi} ~\Bigg/~
  {\int_0^{\pi/2}\frac{\sin^2\phi}{\sqrt{(c\cos^2\phi+\sin^2\phi)(c\sin^2\phi+\cos^2\phi)}}\d\phi}\,.
\label{e:Gammadef}
\end{gather}
\ese
The integrals in~\eqref{e:Gammadef} can be evaluated explicitly,
yielding
\be
s_3^+ = c\frac{c E(m_c) - K(m_c)}{c K(m_c) - \Pi(\alpha^2_c,m_c)}\,,
\qquad
\alpha^2_c = 1-1/c\,,\quad 
m_c = 1 - 1/c^2\,.
\label{e:s3+}
\ee
The behavior of $\l_2$ and $\l_3$ within the 2 transition regions, 
which is not specified in~\eqref{e:solution_c>1b}
(and which was not given in \cite{blochkodama}), 
is discussed in detail in step~2 below.
We point out that all the characteristic speeds above can be obtained as limiting cases of the explicit formulae for the characteristic speeds 
for the genus-one Whitham equations described in section~\ref{s:elliptic_whitham}, see Eq.~\eqref{e:s1s4}.

\paragraph{Step~2.} \label{sec:case2}

Note that the IC \eqref{e:Case1_IC_g1} possess a spatial reflection symmetry:
$\l_1(x,0) = - \l_4(-x,0)$ and 
$\l_2(x,0) = - \l_3(-x,0)$.
Moreover, one can check that the structure of the modulation equations guarantees that this symmetry is preserved by the time evolution.
Because of this, it is sufficient to discuss the solution for $x<0$.
We begin by noting that, since the values of all four invariants are known at the edges of the ``Whitham zone'',
\eqref{e:mdef} allows us to obtain the value of $m$ at the edges of the DSW.
Specifically:
(i) At $x = -s_3^-t$, we have $\l_3 = \l_4 = 2(c+1)$, $\l_2 = -2(c-1)$ and $\l_1 = -2(c+1)$, implying $m=1$.
Therefore, the leading edge of the DSW is a solitary wave.
(ii) At $x = -s_3^+t$, we have $\l_4 = 2(c+1)$, $\l_2 = -2(c-1)$ and $\l_1 = -2(c+1)$, as before,
but now $\l_3 = 2(c-1)$, implying $m = m_c = 1-1/c^2$ [as anticipated in~\eqref{e:s3+}].
Therefore, the trailing edge of the DSW is not a harmonic wave.
This also means that in the central region, $-s_3^+t < x < s_2^-t$, the solution is an exact periodic solution of the Toda lattice (indeed, it is a binary oscillation \cite{VDO}).

Next, note that, since the values of $\l_1$, $\l_2$ and $\l_4$ are constant in the ``Whitham zone'' $- s_3^- t < x < -s_3^+ t$, 
\eqref{e:mdef} allows us to express $\l_3$ as a function of $m$ as
\bse
\[
\l_3(m) = 2 (c+1) \frac{1-c(1-m)}{1+c(1-m)}\,.
\label{e:lambda3case1}
\]
Recall that in Sec.~\ref{s:elliptic_whitham} we simplified the abstract formulation of the speeds
given in Eq.~\eqref{e:g1speeds} to the more practical formulae given in Eq.~\eqref{e:s1s4}.
Combining Eqs.~\eqref{e:s1s4} and ~\eqref{e:lambda3case1}
%Then, in light of the calculations of the preceding sections as well as \eqref{e:lambda3case1} 
and the values of 
$\l_1$, $\l_2$ and $\l_4$, we have $s_3(\fbf\l) = s_3(m)$, with 
\[
s_3(m) = \frac{c(c+1)}{1+c(1-m)}
  \frac{(1+c(1-m)) E(m) - (c+1)(1-m) K(m)}{(c+1) K(m) - (1+c(1-m))\Pi \left(\left.\frac{cm}{c+1}\right|m\right)}\,.
\label{e:s3m_case1}
\]
Note that, one can recover $s_3^{\pm}$ given in Eqs.~\eqref{e:s3-} and \eqref{e:s3+}
by taking the limits $m\rightarrow 1$ and $m\rightarrow m_c$, respectively, in
Eq.~\eqref{e:s3m_case1}.

Next, looking for a self-similar solution of the genus-one Whitham equation~\eqref{e:TodaWhitham}, namely
$\l_3 = \l_3(x/t)$, 
we obtain an equation that determines $x/t$ as a function of $\l_3$ or equivalently $m$:
\[
x/t = - s_3(m)\,,
\label{e:xi_case1}
\]
\ese
with $s_3(m)$ as above.
One can now combine \eqref{e:lambda3case1} and~\eqref{e:xi_case1} to obtain $\l_3$ as a function of $x$,
parametrically in terms of~$m$,
for any fixed~$t$.
Figure~\ref{f:1}(d) shows the value of the Riemann invariants at $t=1$.
These speeds were validated by careful comparison with direct numerical simulations
(see the following section), showing excellent agreement.

\paragraph{Step~3.}

By combining the results of the previous step with the expressions for $a_\mathrm{min}$, $a_\mathrm{max}$ and $a_\mathrm{mean}$ in the preceding section,
we can plot the envelope and mean of the oscillations for any fixed $t$ as a function of $x$, as parametrized by 
the value of $m$.
These values were also validated by careful comparison with direct numerical simulations (in the following section),
again showing excellent agreement.

\paragraph{Step~4.}

The last step is accomplished using a similar method as for step~3, but now combining that with the expression
for the periodic traveling wave solutions of the Toda lattice presented in section~\ref{s:elliptic}.
Namely, for any fixed value of $t$, one sets $n\equiv x$ as given by~\eqref{e:xi_case1} and uses the resulting 
expression in \eqref{e:aelliptic} and~\eqref{e:belliptic}.
Note that the easiest way to compare the analytical predictions with the results of direct numerical simulations
is to compute $a_n(t)$, %since that's $b_{n,\mathrm{T}}$, 
whose expression [\eqref{e:belliptic}] is 
simpler than $b_n$ (in other words, we will compare the velocity, since $\dot{y} = -2a_n(t)$).
The resulting expressions were validated by careful comparison with direct numerical simulations,
once more showing excellent agreement.

\subsection{Case~2: $0<c<1$}

The implementation of the various steps for case~2 is quite similar to that for case~1, 
so, to avoid unnecessary repetition, we will keep the presentation for this case more concise, 
limiting ourselves to pointing out the differences between the two cases
and giving the relevant formulae.

\begin{figure}[t]
\centerline{
   \begin{tabular}{@{}p{0.33\linewidth}@{}p{0.33\linewidth}@{}p{0.33\linewidth}@{}}
     \rlap{\hspace*{5pt}\raisebox{\dimexpr\ht1-.1\baselineskip}{\bf (a)}}
 \includegraphics[height=4cm]{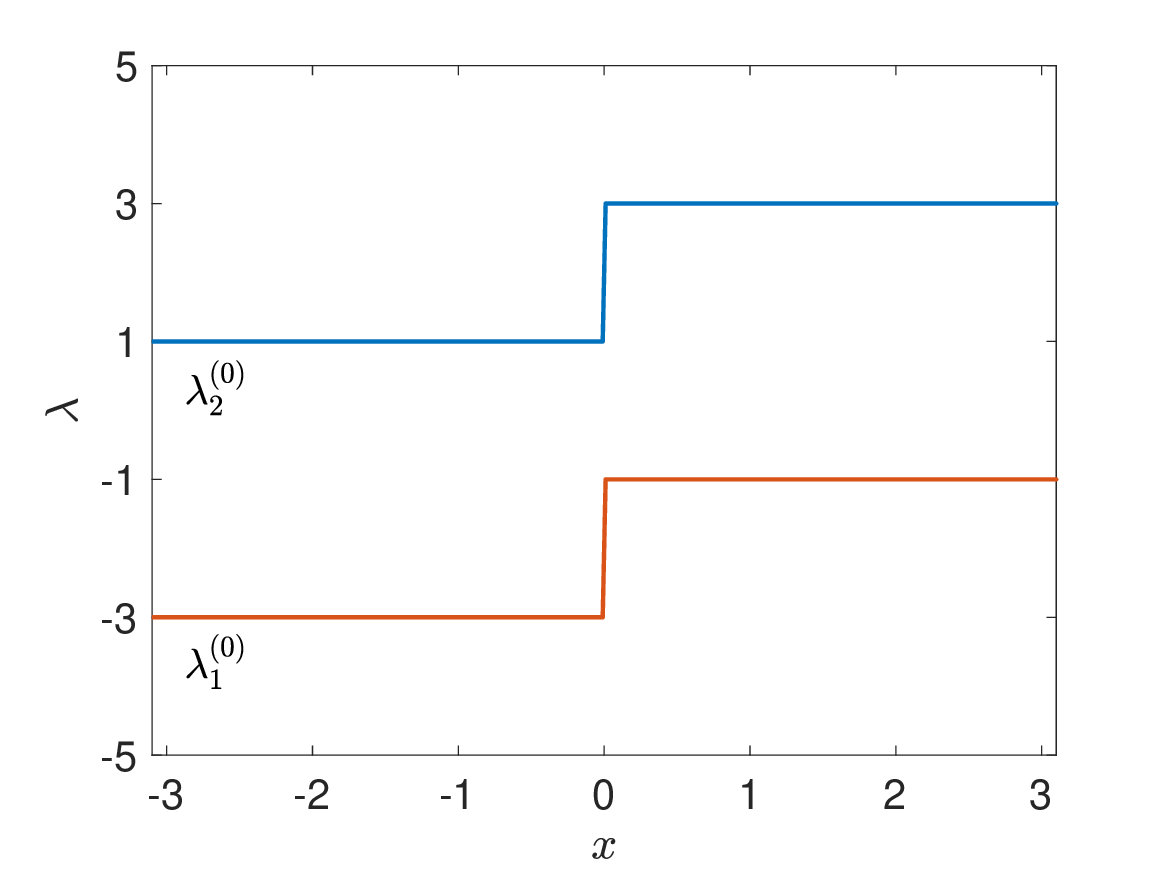} &
      \rlap{\hspace*{5pt}\raisebox{\dimexpr\ht1-.1\baselineskip}{\bf (b)}}
 \includegraphics[height=4cm]{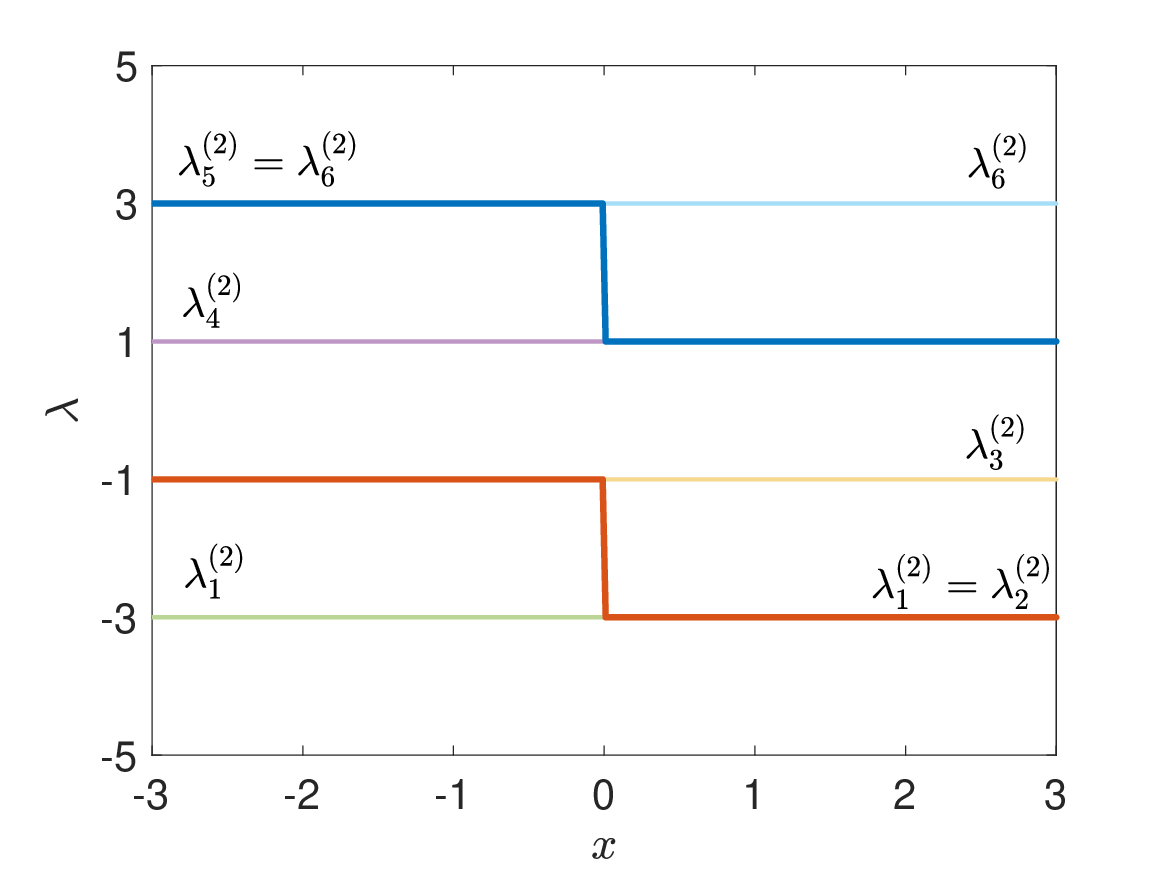} &
  \rlap{\hspace*{5pt}\raisebox{\dimexpr\ht1-.1\baselineskip}{\bf (c)}}
 \includegraphics[height=4cm]{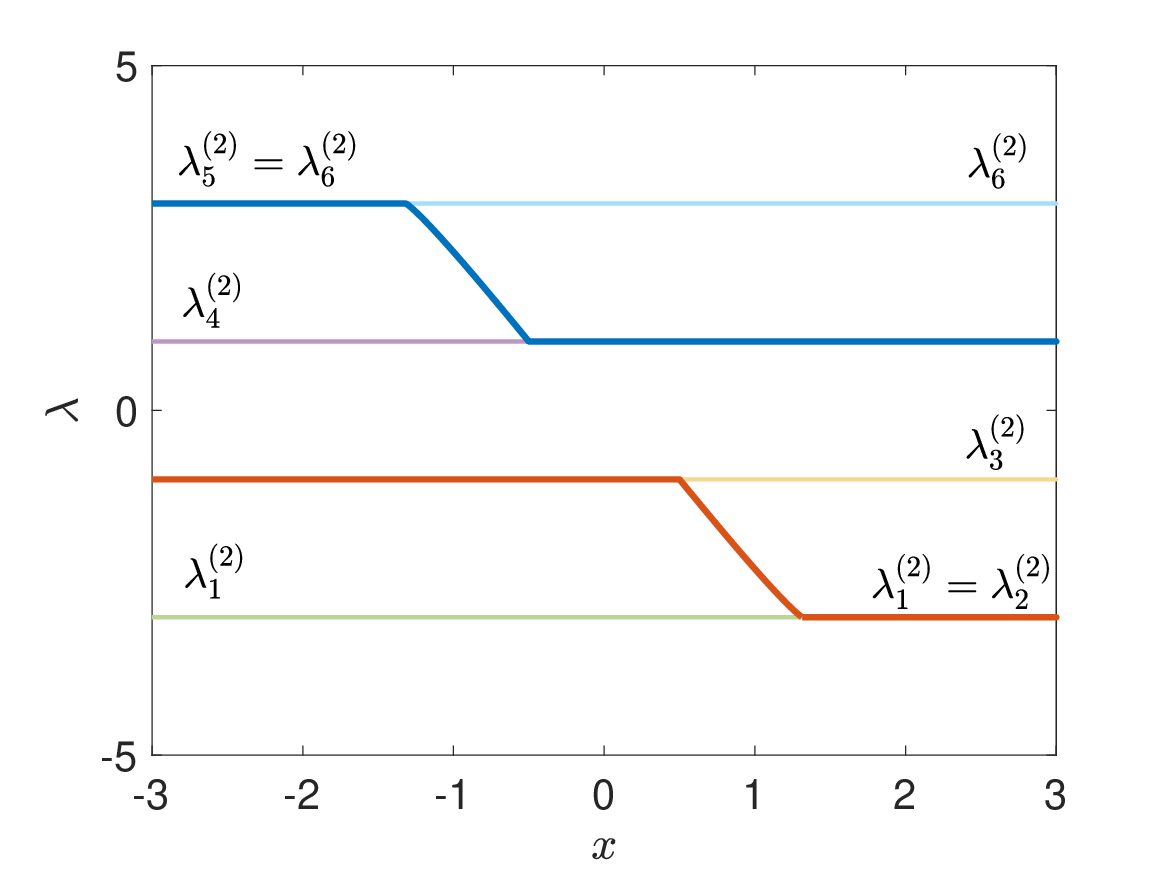} 
  \end{tabular}
  }
  \centerline{
   \begin{tabular}{@{}p{0.33\linewidth}@{}p{0.33\linewidth}@{}p{0.33\linewidth}@{}}
     \rlap{\hspace*{5pt}\raisebox{\dimexpr\ht1-.1\baselineskip}{\bf (d)}}
 \includegraphics[height=4cm]{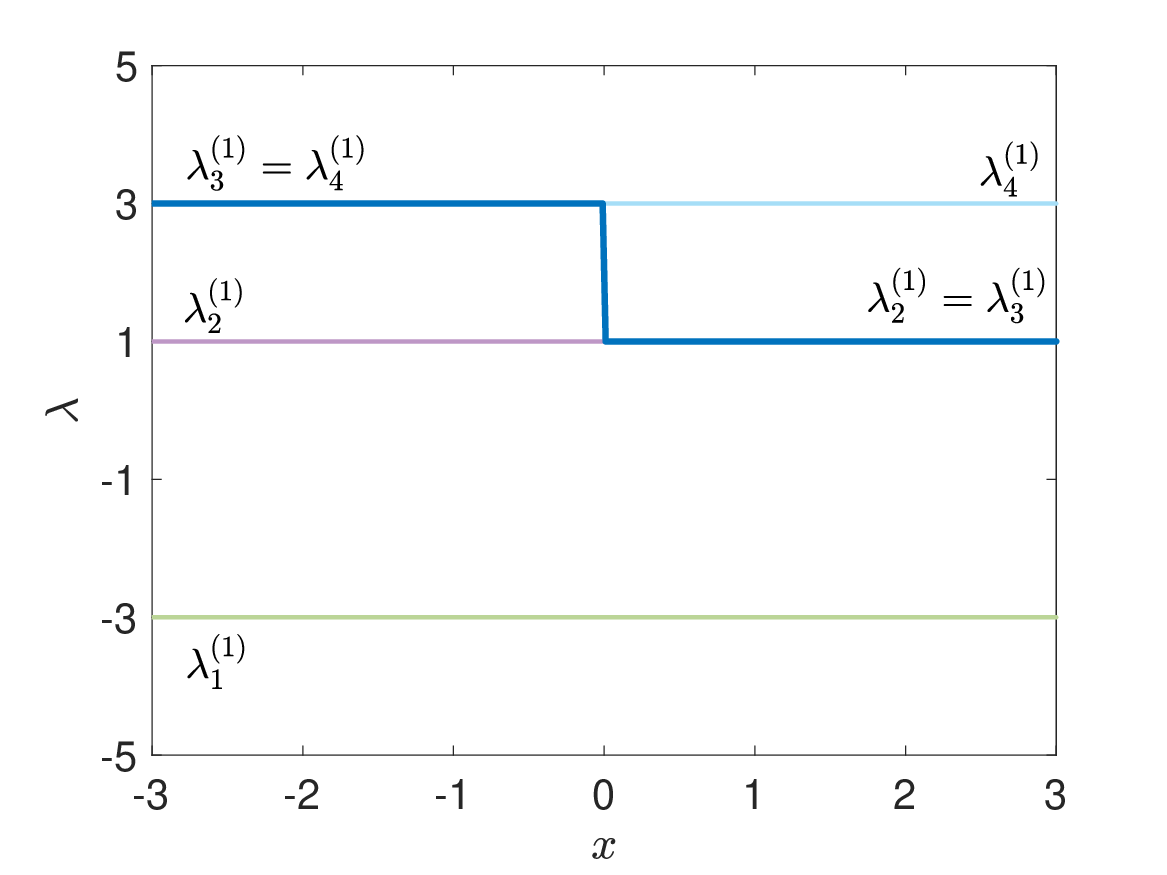} &
      \rlap{\hspace*{5pt}\raisebox{\dimexpr\ht1-.1\baselineskip}{\bf (e)}}
 \includegraphics[height=4cm]{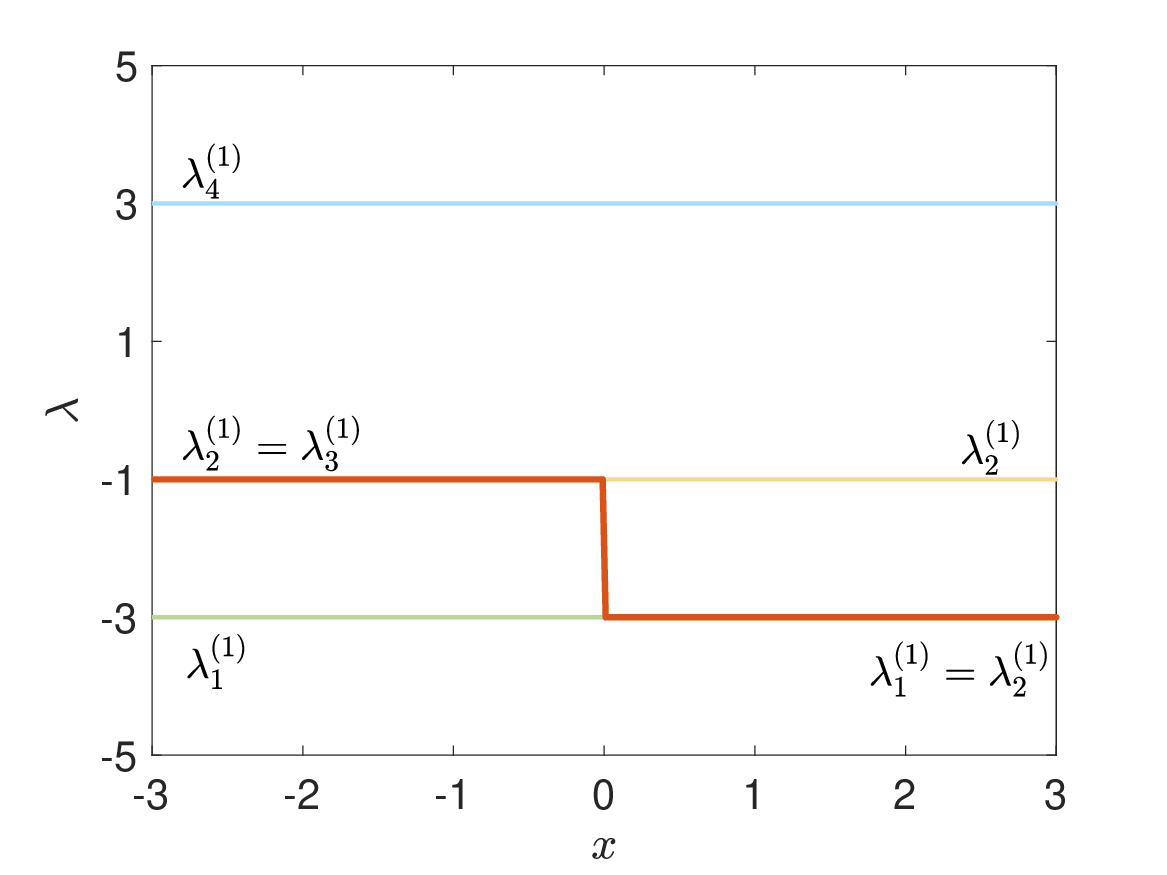} &
  \rlap{\hspace*{5pt}\raisebox{\dimexpr\ht1-.1\baselineskip}{\bf (f)}}
 \includegraphics[height=4cm]{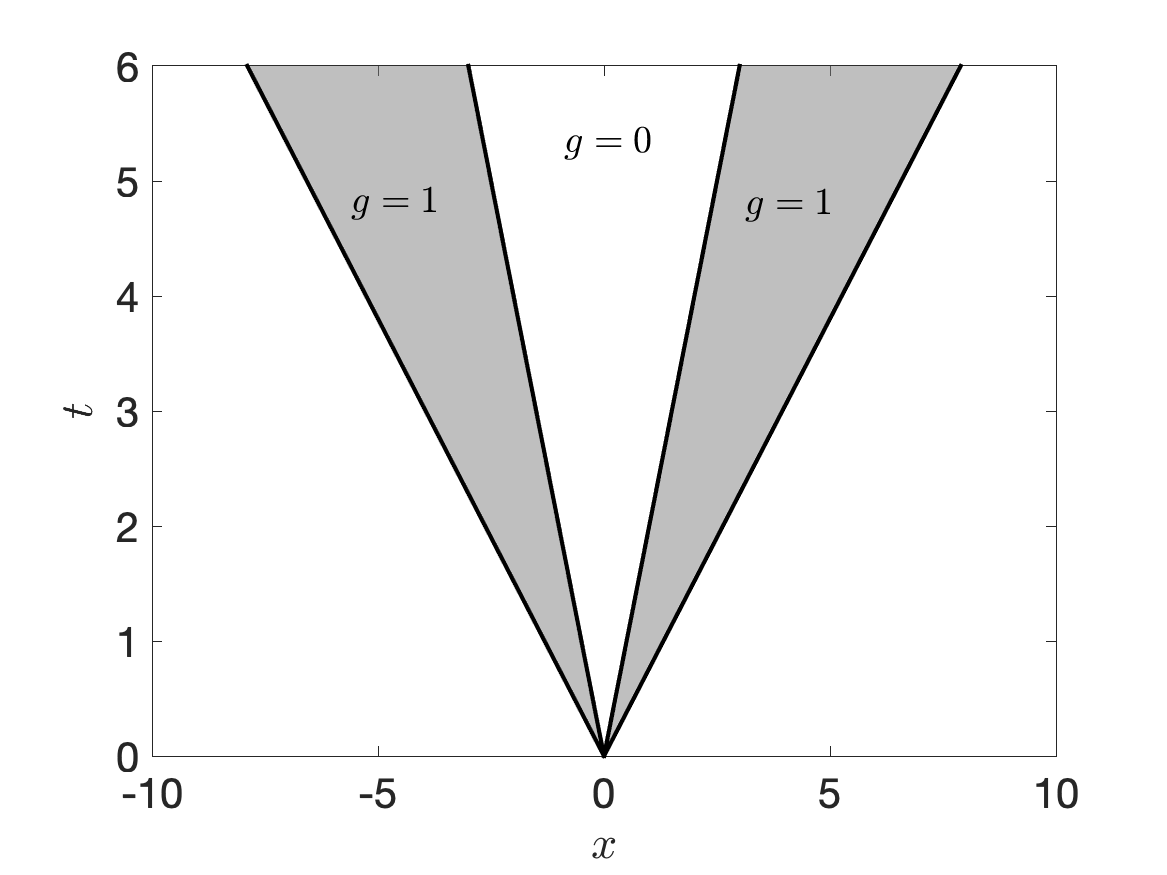} 
  \end{tabular}
  }
\caption{%
\textbf{(a)} ICs for the genus-zero Whitham system %(dark blue: $\l_2$; orange: $\l_1$) 
for case~2 ($c=1/2$).
\textbf{(b)} Regularization of these ICs via the genus-two Whitham equations.
\textbf{(c)} Snapshot of the profile of the Riemann-invariants at $t=1$.  
\textbf{(d)} Effective genus-one regularization for $x<0$.
\textbf{(e)} Effective genus-one regularization for $x>0$.
\textbf{(f)} Bifurcation plot showing the genus-one regions (in light gray) and genus-zero regions (in white) in the $xt$-plane.
}
\label{f:case2}
\end{figure}

Recall that, when $0<c<1$, the ICs for $a_n$ and $b_n$, still given in~\eqref{e:ICblochkodama} are regularized by genus-two data.
The ICs for the genus-zero Whitham equations, still given by~\eqref{e:ICblochkodama_g0},
also give rise to a shock.
In this case, however, to fully regularize the problem it is necessary to embed the ICs as a degenerate case of the
genus-two Whitham equations. The corresponding ICs are
\begin{eqnarray*}
\lambda_1(x,0) = -2 (c + 1),  \quad 
\lambda_2(x,0) = \begin{cases} 2 (c - 1), & x<0, \\ -2 (c+ 1), &x>0, \end{cases} \quad  
\lambda_3(x,0) =  2 (c - 1), 
\\
\lambda_4(x,0) =  -2 (c - 1), \quad 
\lambda_5(x,0) =  \begin{cases} 2 (c + 1), & x<0, \\  -2 (c - 1), &x>0, \end{cases} \quad   
\lambda_6(x,0) =  2(c + 1).
\end{eqnarray*}
The upshot is that the two DSWs are located in the regions 
$-s_5^- t < x < -s_5^+ t$ and $-s_2^- t < x < -s_2^+ t$, 
where 
(it is trivial to show that $s_5^->s_5^+$ for all $0<c<1$)
\be
s_5^- = - s_2^+ = \frac{\sqrt{c(c+1)}}{\log(\sqrt{c}+\sqrt{c+1})}\,,\qquad
s_5^+ = - s_2^- = 1 - c\,. \label{e:s5}
\ee

In general, one cannot obtain explicit expressions for the Riemann invariants 
of the genus-two Whitham equations in the Whitham zones, even in parametric form.
Similarly, a quantitative comparison between the predictions of the genus-two Whitham theory
and the actual behavior of the solutions of the Toda lattice would require evaluation of the expressions
of the finite-genus solutions of the Toda lattice in terms of theta function \cite{Teschl}. 
Importantly, however, even though one needs to employ the genus-two Whitham equations
in order to regularize the ICs for case~2 for all $x\in\Real$, 
it is, in fact, possible to compute all the necessary quantities using only the genus-one Whitham equations,
as long as one uses two different genus-one reductions to study negative and positive values of~$x$. 

The key observation that makes this possible is that, even though the regularization of the ICs involves the genus-two Whitham equations, no genus-two region is ever produced by the time evolution. 
Accordingly, only four Riemann invariants are distinct for $x<0$,
and the same is true for $x>0$.
The two set of four invariants are not the same, which requires one to use six invariants in order to regularize the ICs for all~$x$.
However, as long as we are interested in only studing positive or negative values of~$x$, 
we can use two different regularizations in parallel: 
a first one in order to compute the solution for $x<0$ and a second one to compute the solution for $x>0$.
Specifically, denoting by $\l_j\O{g}$ the Riemann invariants of the genus-$g$ Whitham equations,
and referring to Fig.~7 in \cite{blochkodama}, we have the following:

%\begin{description}
%\item[4.2.1.]
To study the region $x<0$, we set
\be
\l_1\O1 = \l_1\O2,\quad
\l_2\O1 = \l_4\O2,\quad
\l_3\O1 = \l_5\O2,\quad
\l_4\O1 = \l_6\O2,
\ee
thereby neglecting $\l_2\O2$ and $\l_3\O2$, whose values coincide for all $x<0$.
The corresponding ICs are then
\be
\l_1\O1 = -2(1+c),\qquad
\l_2\O1 = 2(1-c),\qquad
\l_3\O1 = \begin{cases} 2(1+c), &x<0,\\ 2(1-c), &x>0, \end{cases}\qquad
\l_4\O1 = 2(1+c).
\ee
%\item[4.2.2.] 
To study the region $x>0$, we set
\be
\l_1\O1 = \l_1\O2,\quad
\l_2\O1 = \l_2\O2,\quad
\l_3\O1 = \l_3\O2,\quad
\l_4\O1 = \l_6\O2,
\ee
thereby neglecting $\l_4\O2$ and $\l_5\O2$, whose values coincide for all $x<0$.
The corresponding ICs are
\be
\l_1\O1 = -2(1+c),\qquad
\l_2\O1 = \begin{cases} -2(1-c), &x<0,\\ -2(1+c), &x>0, \end{cases}\qquad
\l_3\O1 = -2(1-c),\qquad
\l_4\O1 = 2(1+c).
\ee
%\end{description}

With the above setup, we can use the framework developed in the previous section to study the dynamics.
Similarly to case~1, the Riemann invariants again satisfy the symmetry $\l_{j}(x,t) = - \l_{7-j}(-x,t)$ for all $x\in\Real$, $j=1,\dots,6$.
Therefore it is sufficient to just present the results for $x<0$.
Using the effective genus-one regularization for $x<0$ presented above, one finds that,
in the Whitham modulation zone $-s_5^- t < x < - s_5^+ t$, 
$\l_3\O1(m)$ is still given by~\eqref{e:lambda3case1}.
Moreover, the corresponding characteristic speed is
also still given by~\eqref{e:s3m_case1}.
Similarly to before, one can then look for a self-similar solution of the Whitham equations, i.e., $\l_5 = \l_5(x/t)$,
obtaining the analogue of~\eqref{e:xi_case1} $x/t = - s_5(m)$.
The remaining steps in the analysis are completely analogous to those in case~1.

\section{Direct numerical simulations and comparison with the theory}
\label{s:simulations}

The previous section describes how to produce an analytical prediction of
the leading and trailing edge speed, the maximum, minimum and mean value
of the solution, and the actual spatio-temporal profile itself. We now compare
these predictions against numerical simulations.

We simulate Eq.~\eqref{e:Toda_d2ydt2} with $M=k=1$ using the symplectic Verlet scheme
with a time step size of $\Delta t = 0.001$, \cite{hairer2013geometric}.
While we are cognizant of important recent work illustrating
the effect of such integrators on imposing an effective periodic 
drive, and ultimately leading to a breakdown of integrability
over (very) long time simulations~\cite{danieli2023dynamical},
such effects are not relevant over the timescales
considered herein.
The simulation is initialized with the shock initial data
given by Eq.~\eqref{e:TodashockIC}.  In the simulations shown
here we selected $N=8000$ nodes, such that the leftmost index is $N_1 = -4000$
and the rightmost index is $N_2 = 3999$. We employ free boundary
conditions, that is $y_{N_1} = y_{N_1+1}$ and $y_{N_2} = y_{N_2-1}$.
For the simulations considered, the lattice is sufficiently large 
to avoid interactions between the excitations and the boundaries.
We perform simulations to cover representative examples
of the two cases presented above, namely $c = 1.5$ for case 1
and $c = 0.5$ for case 2.

\subsection{Case 1: $c>1$}

\begin{figure}[t!]
\centerline{
   \begin{tabular}{@{}p{0.5\linewidth}@{}p{0.5\linewidth}@{}}
     \rlap{\hspace*{5pt}\raisebox{\dimexpr\ht1-.1\baselineskip}{\bf (a)}}
 \includegraphics[height=6.525cm]{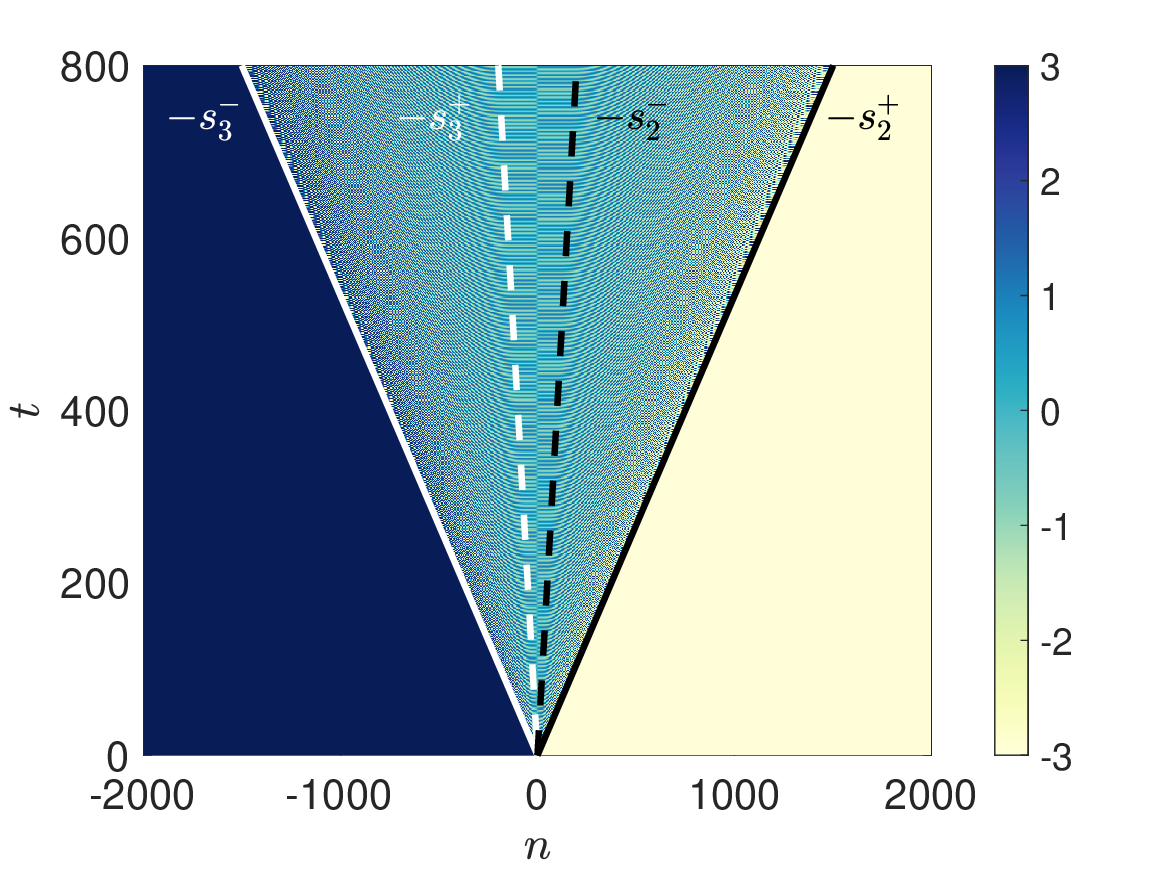} &
  \rlap{\hspace*{5pt}\raisebox{\dimexpr\ht1-.1\baselineskip}{\bf (b)}}
 \includegraphics[height=6.525cm]{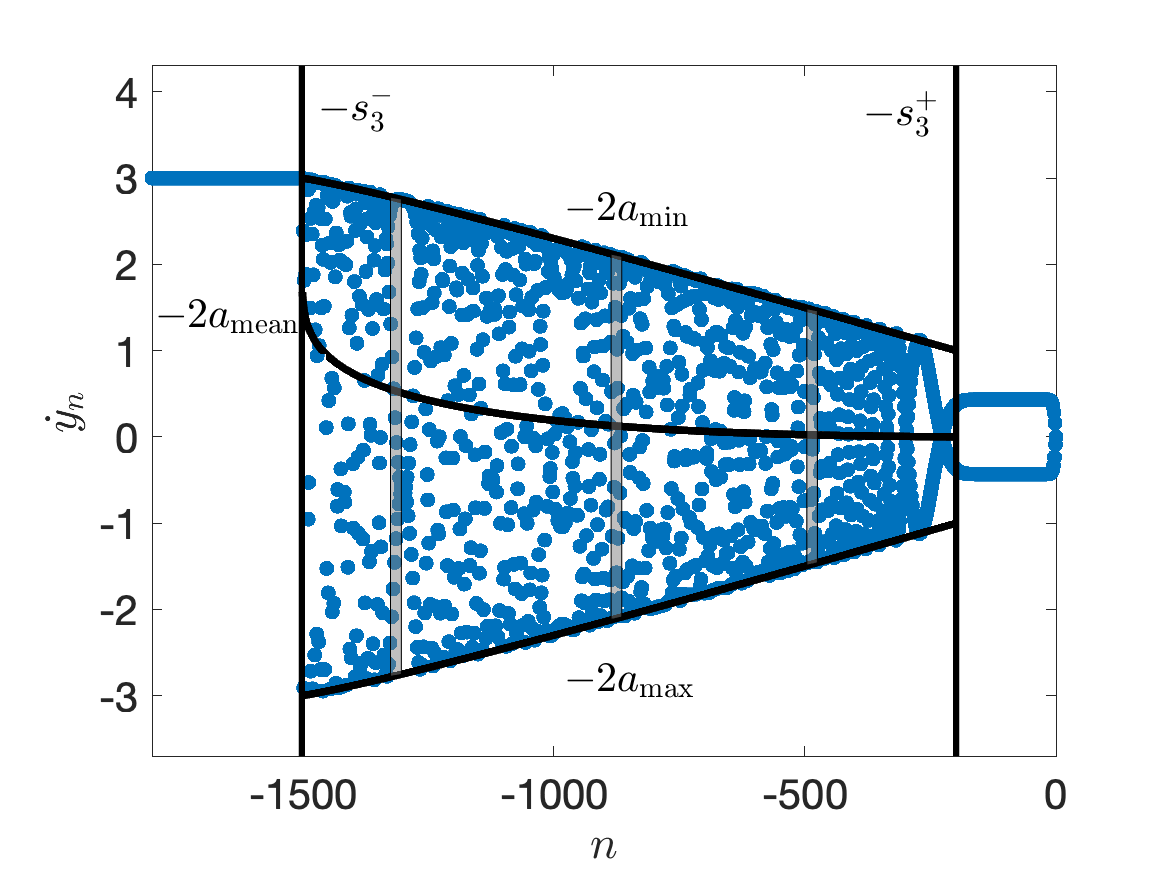} 
  \end{tabular}
  }
\centerline{
   \begin{tabular}{@{}p{0.33\linewidth}@{}p{0.33\linewidth}@{}p{0.33\linewidth}@{} }
     \rlap{\hspace*{5pt}\raisebox{\dimexpr\ht1-.1\baselineskip}{\bf (c)}}
 \includegraphics[height=4.3cm]{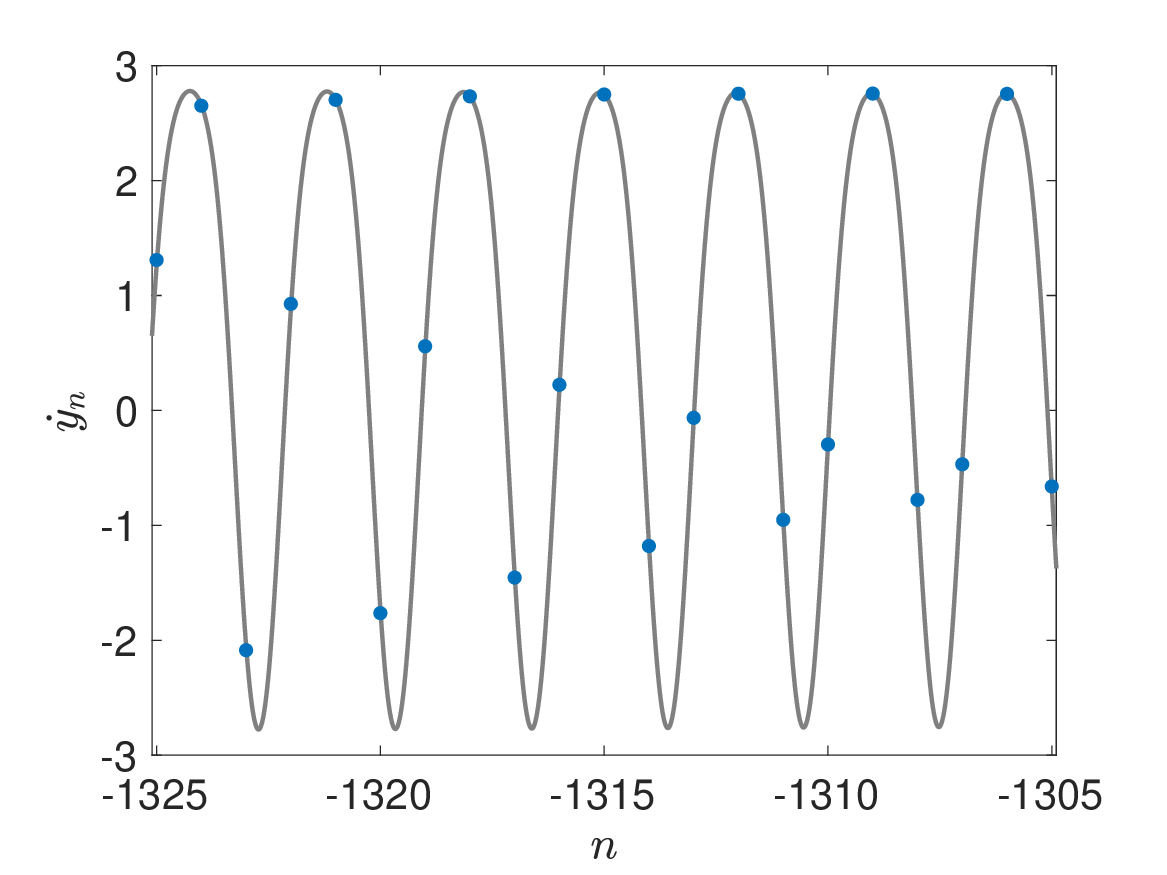} &
  \rlap{\hspace*{5pt}\raisebox{\dimexpr\ht1-.1\baselineskip}{\bf (d)}}
 \includegraphics[height=4.3cm]{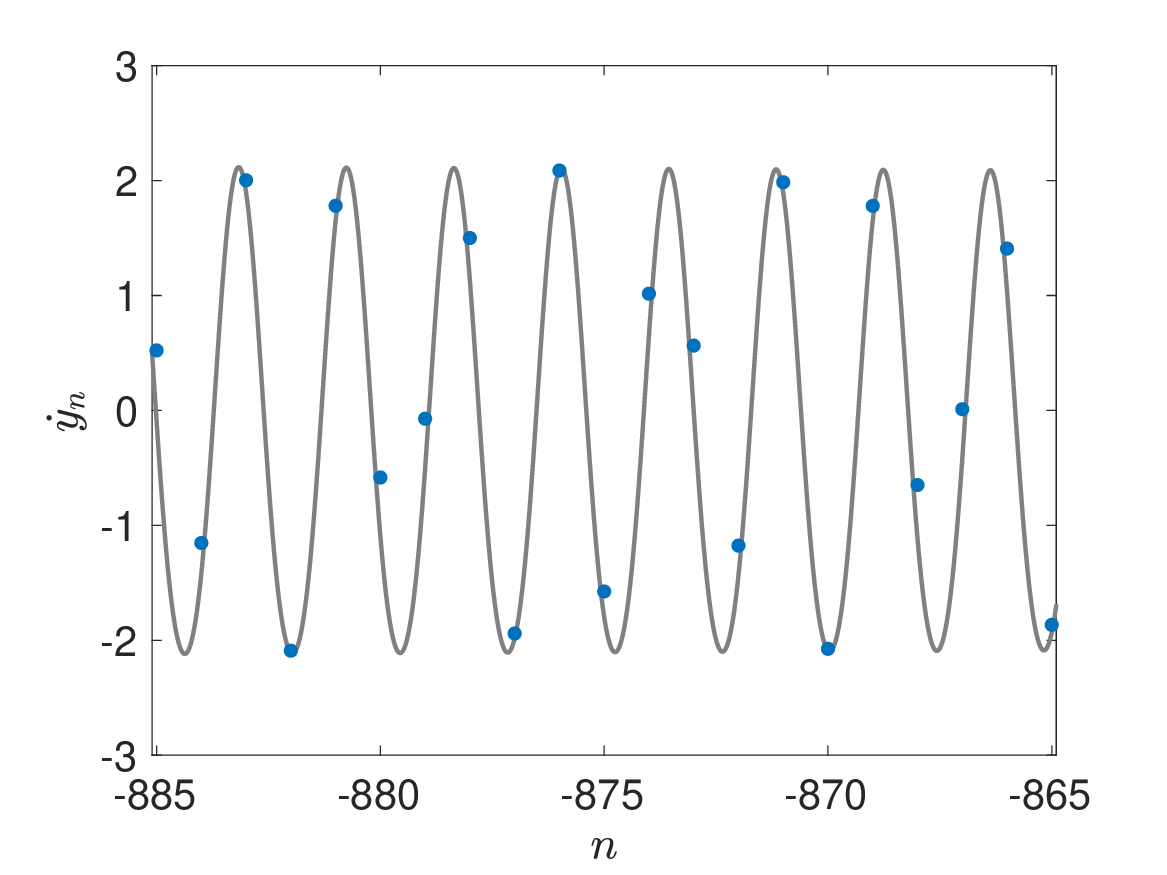} &
   \rlap{\hspace*{5pt}\raisebox{\dimexpr\ht1-.1\baselineskip}{\bf (e)}}
 \includegraphics[height=4.3cm]{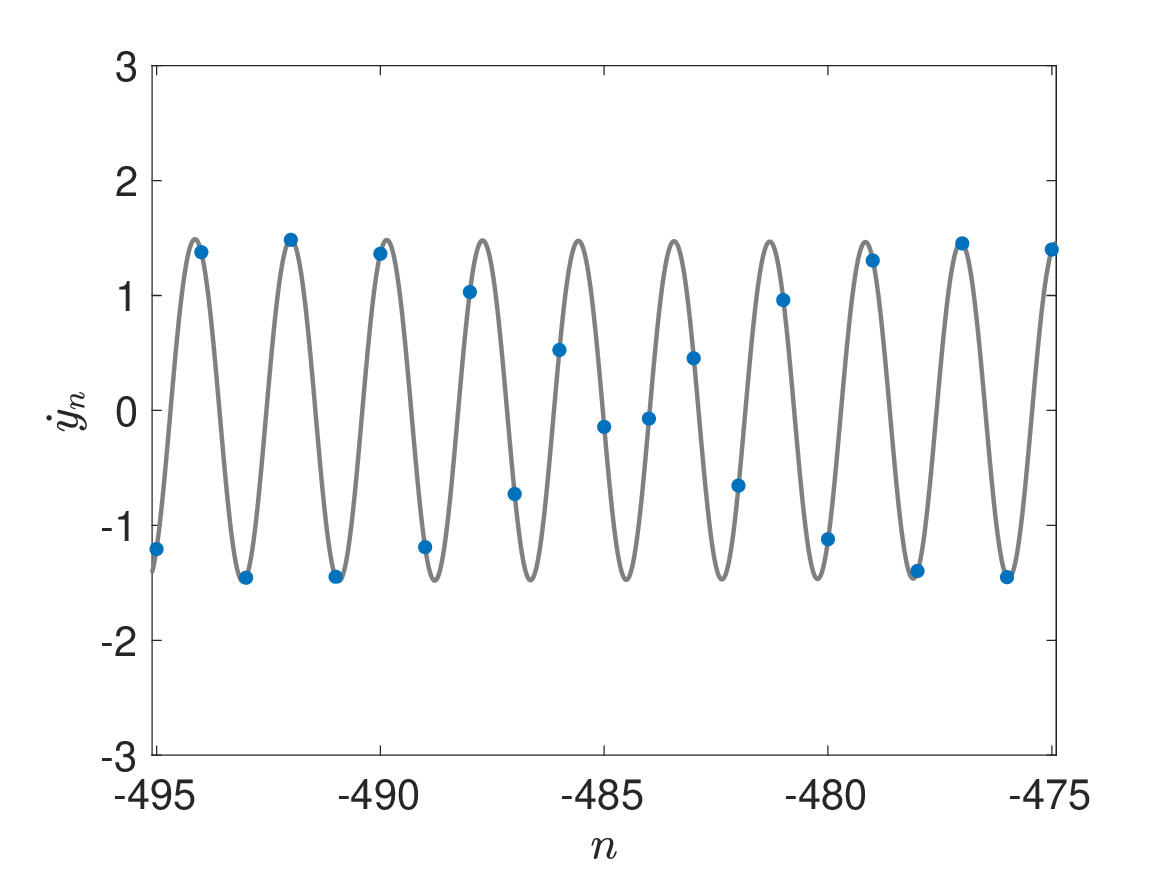} 
  \end{tabular}
  }
    \caption{ \textbf{(a)} Intensity plot of the DSW of Eq.~\eqref{e:Toda_d2ydt2} with
    the shock initial data, Eq.~\eqref{e:TodashockIC}, with $c=1.5$. The solid and dashed curves
    represent the straight space-time lines associated with the trailing and leading edge speed predictions, respectively. Color intensity corresponds to the
    velocity. \textbf{(b)} Zoom of velocity profile of the solution shown in (a) at time $t=800$. The vertical solid
    lines are the theoretical predictions of the trailing and leading edges. The curved line is the mean value prediction,
    and the sloped lines represent the envelope prediction of the DSW. \textbf{(c)} Zoom of DSW shown in (b)
    at the first gray shaded region (blue markers). The gray solid curve is the theoretical prediction.
    \textbf{(d)} Same as (c), but the zoom corresponds to the middle gray shaded region of (b). \textbf{(e)} Same as (c), but the zoom corresponds to the last gray shaded region of (b).
    } 
    \label{fig:c1p5}
\end{figure}

We first compare the formulae of Section~\ref{sec:case1} with numerics. A 2D density plot
of a simulation initialized by Eq.~\eqref{e:TodashockIC} with $c=1.5$
is shown in Fig.~\ref{fig:c1p5}(a). This is a standard way to present the
evolution of a DSW
\cite{PRE98p052220,CMP382p1496,PRE98p042211,SIREV60p888,PRE99p022215,PRL116v043902,CPAM70p2300}.
In the figure, color intensity corresponds
to the velocity, $\dot{y}_n$.  Note, while the computational window
consisted of $N=8000$ nodes, only $4000$ are shown for clarity purposes.
Two counter-propagating and expanding waveforms can be identified,
which are the DSWs.  In the first region (where the lattice index
is negative) the leading and trailing edge of the DSW are predicted
by the $-s_3^-$ and $-s_3^+$ formulae, respectively, which are
defined in Eq.~\eqref{e:s3-}.  %and ~\eqref{e:s3+}, respectively.
For the calculation of $s_3^+$, we use the symbolic formula \eqref{e:s3+}
instead of numerically evaluating Eq.~\eqref{e:Gammadef}.
This requires the computation of the
complete elliptic integrals of the first ($E(m)$), second ($K(m))$,
and third ($\Pi(n,m)$) kinds. These
 are standard in most computational libraries
(for example in Matlab, they are computed as \texttt{[K,E]=ellipke(m)}
and \texttt{$\Pi(n,m)$= ellipticPi(n, m)}  ).

%\paragraph{Step~1.}
In Fig.~\ref{fig:c1p5}(a),
a plot of $t=-n/s_3^-$ (solid white line) and $t=-n/s_3^+$ (dashed white
line) are shown, which compare very well to the leading and trailing
edge of the DSW in the first region. A similar plot of $-n/s_2^-$ (solid black line) and $-n/s_2^+$ (dashed black
line) are also shown, which compare very well to the leading and trailing
edge of the DSW in the second region. 
These two regions correspond to
the genus-one regions of the Whitham equations, see also the gray regions
of Fig.~\ref{f:1}(c).
In the genus-zero region
(the region between $-n/s_3^+$ and $-n/s_2^-$, see also the central white region
of Fig.~\ref{f:1}(c))
there are 
binary
oscillations with non-zero amplitude \cite{VDO}.
Note that with $m=m_c$ the wavelength of oscillation
in Eq.~\eqref{e:belliptic} is 2, which corresponds
to a binary oscillation. 
The comparison shown in Fig.~\ref{fig:c1p5}(a) corresponds to step 1 of
Sec.~\ref{s:todadsw}. Note that step 2 of
Sec.~\ref{s:todadsw} does not involve any comparison
to numerical simulation, and hence there is no corresponding figure here,
and we move onto step 3.

In Fig.~\ref{fig:c1p5}(b) a zoom of the velocity profile of the DSW in the first Whitham zone  at time $t=800$ is shown. The vertical solid lines are the theoretical predictions of the leading and
trailing edges located at the lattice positions $n=-800 s_3^-$ and $n=-800 s_3^+$, respectively. 
The minimum and maximum of the analytical prediction of the velocity
are given by Eq.~\eqref{e:bextreme} and the mean is given via \eqref{e:mean}. 
These formulae are parameterized by $E_1,E_2,E_3,E_4$,  which are given
in terms of the Riemann invariants via the formula $2 E_j = \lambda_{j}, j=1,2,3,4$.
In the first Whitham
zone (i.e., the DSW for $n<0$ in Fig.~\ref{fig:c1p5}(a)), the parameters
$$ E_1 = \frac{\lambda_1}{2} = -(1+c), \quad
E_2 = \frac{\lambda_2}{2} = -(c-1), \quad 
E_4 = \frac{\lambda_4}{2} = (c+1), $$
are constant throughout the region and $E_3(m)$ varies as a function of $m$.
In turn, $m$ is parameterized by $n$ and $t$. In practice, it is easier to define values of $m\in[m_c,1]$ (where $m_c = 1 - 1/c^2$ is the cut-off value
of the parameter $m$) and
then compute $n$ via $ n =  s_3(m) t$
where $t=800$ and $s_3(m)$ is given by Eq.~\eqref{e:s3m_case1}. Thus, for each given value of $m\in[m_c,1]$ we can map the value of  
$$E_3(m)  = \frac{\lambda_3(m)}{2} =   (c+1) \frac{1-c(1-m)}{1+c(1-m)}  $$
to a particular coordinate $(n,t)$, and hence we can map
$a_{\rm max},a_{\rm min}$ and $a_{\rm mean}$ 
to $(n,t)$.
To compare these values to the corresponding values of the velocity,
we simply use the definition $\dot{y}_n = -2a_n(t)$.
In Fig.~\ref{fig:c1p5}(b) the sloped solid black lines represent the envelope prediction of the DSW (i.e. $a_{\rm max}$ and $a_{\rm min}$) as a function
of $n$ with $t=800$ fixed. The curved line is the mean value prediction. The solid black lines and curves of Fig.~\ref{fig:c1p5}(b) correspond to step 3 of Sec.~\ref{s:todadsw}.

%\paragraph{Step~4.}
%In this step,
We now will compare the velocity profile itself to the theoretical
prediction. The process is essentially the same as detailed above,
where for each value of $m\in[m_c,1]$ we obtain $E_1,E_2,E_3,E_4$,
and hence we can compute $a_n(t)$, which is defined
in Eq.~\eqref{e:belliptic}. Note that this requires the computation
of $F(z,m)$, which is the inverse of $\sn(z,m)$. Namely,
$F(z,m)$ is the function such that $z = \sn(F(z,m),m)$, i.e.,
$$F(z,m) = \int_0^z \frac{dt}{\sqrt{(1-t^2)(1-mt^2)}}.$$
This function can be approximated using methods
such as those implemented in \cite{Finverse}.
Panels (c) - (e) of Fig.~\ref{fig:c1p5} show a comparison
of the theoretical prediction, given via $-2a(t)$ (gray curves),
and the numerical solution (blue markers). 
Note that, as usual when comparing the predictions of Whitham theory to the underlying dynamics of the original nonlinear system, one must take into account the presence of a slowly-varying translation offset that is not captured by Whitham theory; see, e.g., \cite{Mark2016}.
In our case, the phase shift is also a slowly varying function of $n,t$
(i.e., the $Z_0(n,t)$ term of Eq.~\eqref{e:Z}).
For the purpose of comparison with numerical simulations, we treat
$Z_0(n,t)$ as a fitting parameter. The blue markers in Fig.~\ref{fig:c1p5}(c) 
correspond to the numerical DSW shown in (b) at the first gray shaded region. 
The lattice indices are $n\in[-1325 , -1305]$, which for
$t=800$ corresponds to $m\in[0.9652,0.9696]$. The gray solid curve is the theoretical prediction
given by $-2 a_n(800)$. For this 
spatial window, we chose $Z_0 = 1.3$ to obtain good agreement between theory and simulation.
Panel (d) is the same as (c), but the zoom corresponds to the middle gray shaded region of (b),
in which case $n\in[-885, -865]$  and $m\in[0.8512, 0.8578]$. In this spatial window
$Z_0 = 1.2$. Panel (e) is also the same as (c), but the zoom corresponds to the last gray shaded region of (b), in which case $n\in[-495 , -475]$, $m\in[0.7024, 0.7120]$ and $Z_0 = 1.1$. 
The above choices are in line with the expectation
of slow variation of $Z_0$. The comparisons shown in panels (c)-(e) correspond to step 4 of Sec.~\ref{s:todadsw}.

\begin{figure}[t!]
\centerline{
   \begin{tabular}{@{}p{0.45\linewidth}@{}p{0.45\linewidth}@{}}
     \rlap{\hspace*{5pt}\raisebox{\dimexpr\ht1-.1\baselineskip}{\bf (a)}}
 \includegraphics[height=5cm]{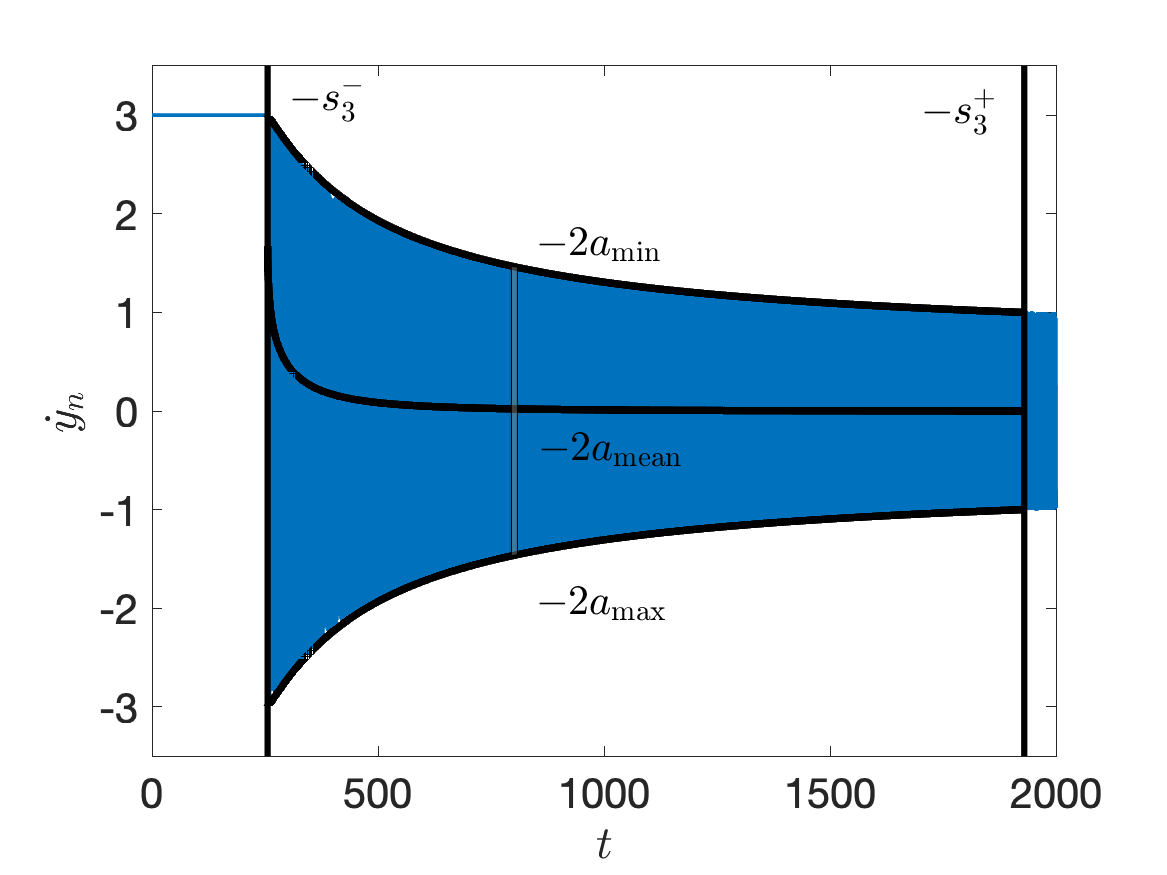} &
  \rlap{\hspace*{5pt}\raisebox{\dimexpr\ht1-.1\baselineskip}{\bf (b)}}
 \includegraphics[height=5cm]{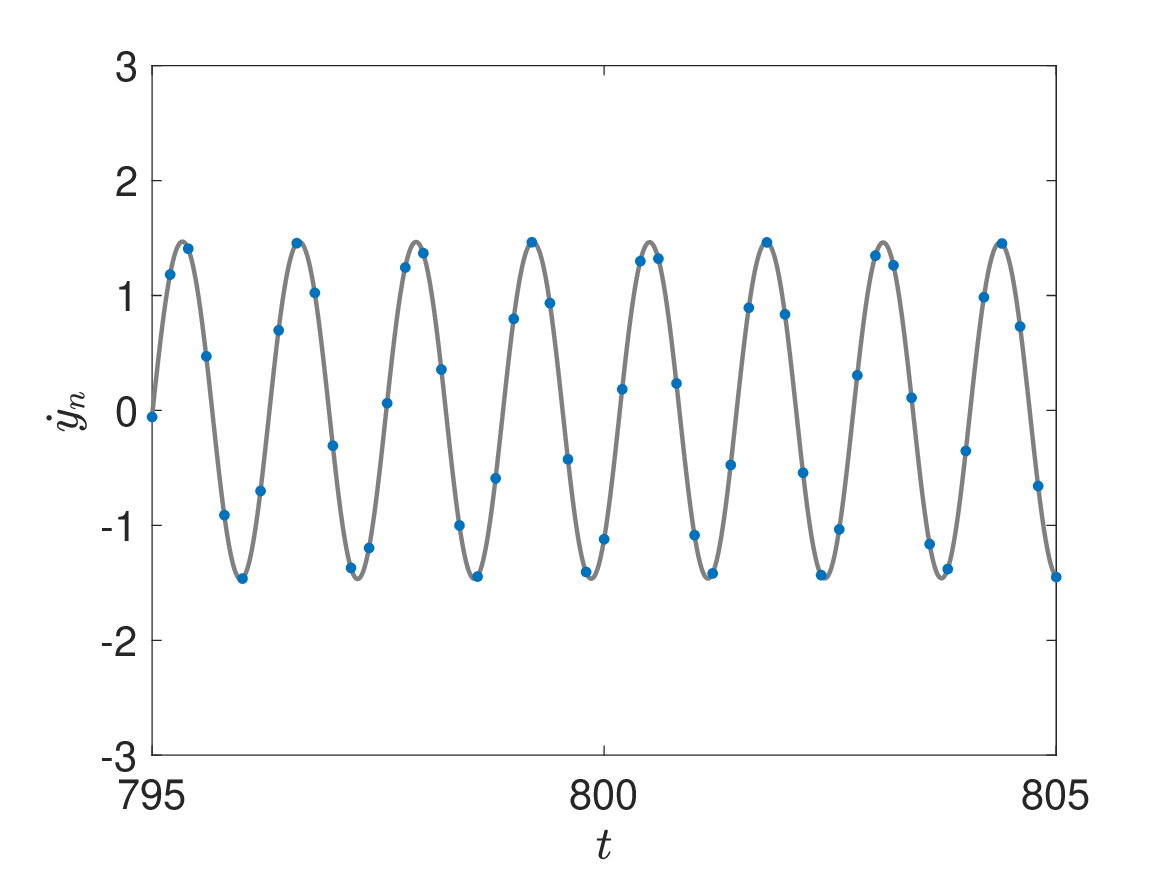} 
  \end{tabular}
  }
\centerline{
   \begin{tabular}{@{}p{0.45\linewidth}@{}p{0.45\linewidth}@{}}
     \rlap{\hspace*{5pt}\raisebox{\dimexpr\ht1-.1\baselineskip}{\bf (c)}}
 \includegraphics[height=5cm]{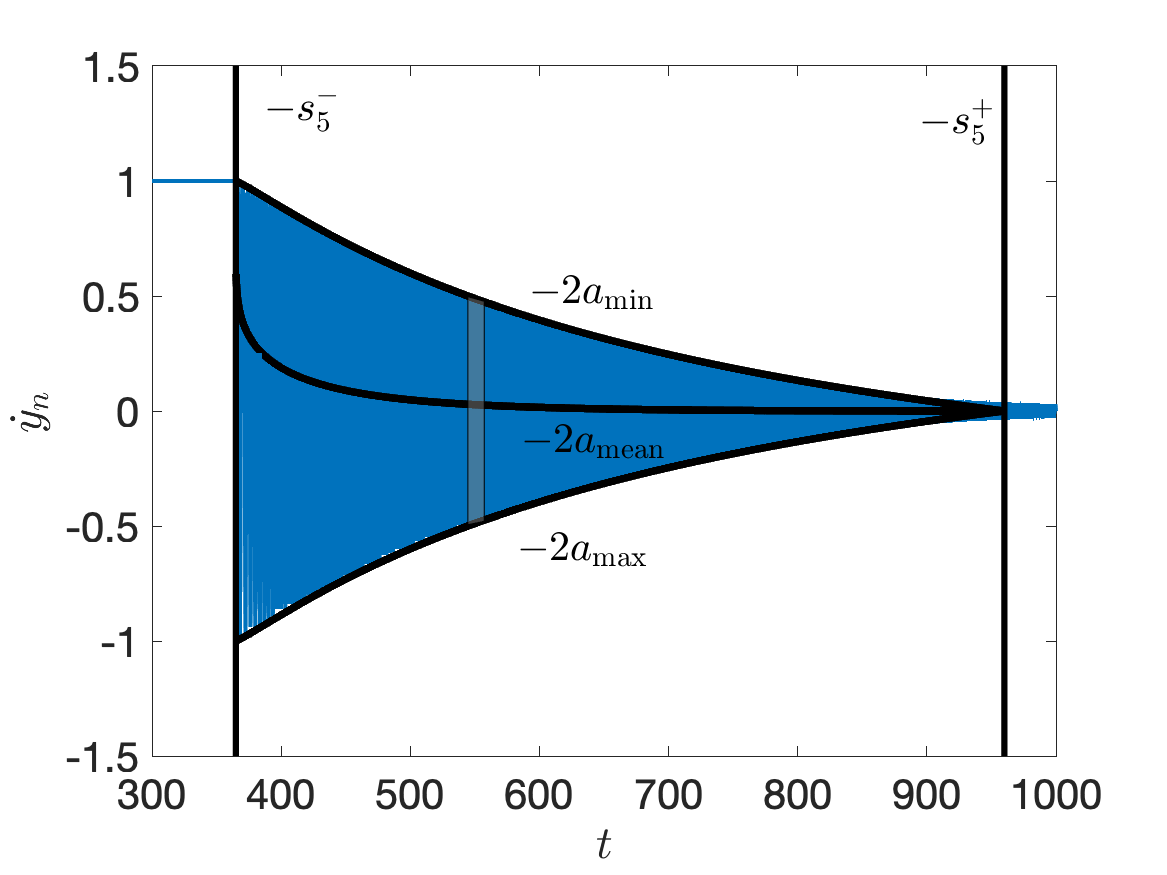} &
  \rlap{\hspace*{5pt}\raisebox{\dimexpr\ht1-.1\baselineskip}{\bf (d)}}
 \includegraphics[height=5cm]{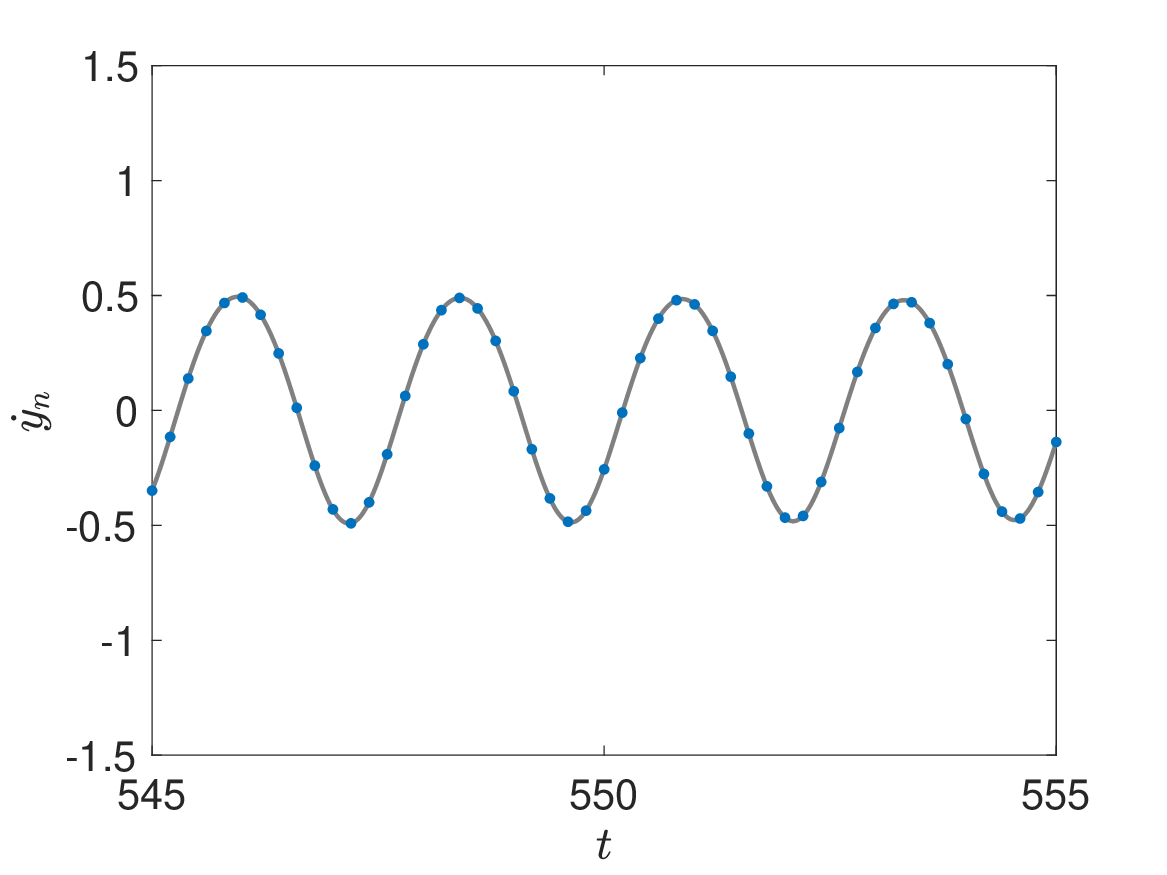} 
  \end{tabular}
  }
    \caption{ \textbf{(a)} Zoom of temporal velocity profile of the solution shown in Fig.~\ref{fig:c1p5}(a) at lattice site $n=-480$. The vertical solid
    lines are the theoretical predictions of the leading edges. The curved line is the mean value prediction,
    and the sloped lines represent the envelope prediction of the DSW. \textbf{(b)} Zoom of DSW shown in (a)
    in the gray shaded region (blue markers). The gray solid curve is the theoretical prediction
    given by $-2 a_{-480}(t)$ where $b_n(t)$ is defined in Eq.~\eqref{e:belliptic}. 
\textbf{(c)} Zoom of temporal velocity profile of the solution shown below in Fig.~\ref{fig:c0p5}(a) at lattice site $n=-480$. The vertical solid
    lines are the theoretical predictions of the leading edges. The curved line is the mean value prediction,
    and the sloped lines represent the envelope prediction of the DSW. \textbf{(d)} Zoom of DSW shown in (c)
    in the gray shaded region (blue markers). The gray solid curve is the theoretical prediction.
    } 
    \label{fig:time_plots}
\end{figure}

Figure~\ref{fig:time_plots} is similar to Fig.~\ref{fig:c1p5}(b)-(e), but the time evolution
is shown and lattice location is fixed to $n=-480$ (which falls within the window
shown in Fig.~\ref{fig:c1p5}(e)). 
Panel (a) of Fig.~\ref{fig:time_plots}
shows a comparison of the time evolution of the numerical simulation (blue lines)
and the predicted maximum, minimum, and mean (black curves). The leading
and trailing edges are also shown, which are the vertical black lines located at the times
$t =480/s_3^-$ and $t = 480/ s_3^+$, respectively. For $t > -s_3^+$, the solution
is time periodic. The time window shown is large relative
to the frequency of oscillation on the microscopic scale, and hence the solution
appears to be a solid blue segment. However, upon zooming into the solution,
one can see the oscillatory structure, as shown in panel (b).
Once again, for the purpose of comparison with numerical simulations, we treat
$Z_0(n,t)$ as a fitting parameter. The blue markers in Fig.~\ref{fig:time_plots}(b) 
correspond to the numerical DSW shown in (a) in the gray shaded region. 
While the time trajectory is continuous, we only plot the solution every
$\Delta t = 0.2$ time units, which allows for easier comparison between
the theory and numerics.
The time window is $t\in[290 , 300]$, which for
$n=-480$ corresponds to $m\in[0.959,0.970]$. The gray solid curve is the theoretical prediction. For this 
temporal window, we chose $Z_0 = 0.68$ to obtain good agreement between theory and simulation.

\subsection{Case 2: $0<c < 1$}

\begin{figure}[t!]
\centerline{
   \begin{tabular}{@{}p{0.5\linewidth}@{}p{0.5\linewidth}@{}}
     \rlap{\hspace*{5pt}\raisebox{\dimexpr\ht1-.1\baselineskip}{\bf (a)}}
 \includegraphics[height=6.525cm]{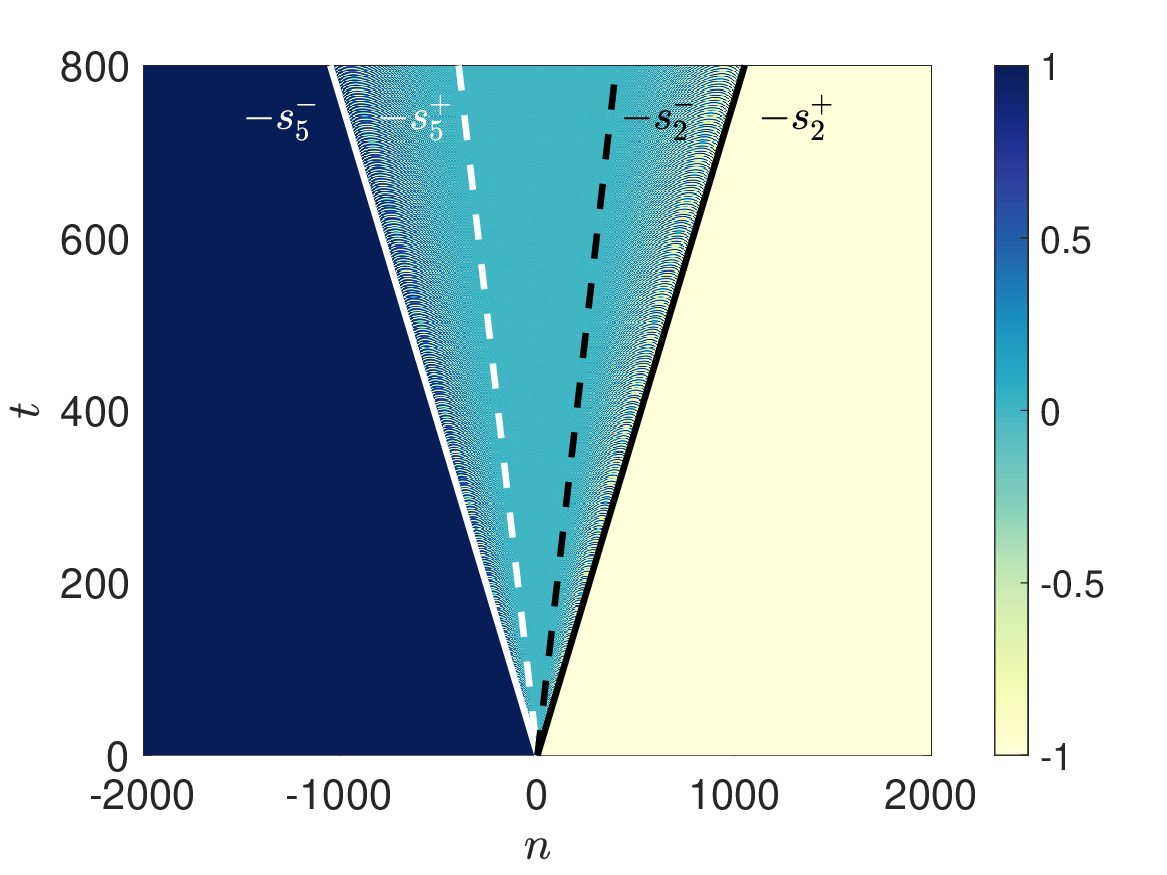} &
  \rlap{\hspace*{5pt}\raisebox{\dimexpr\ht1-.1\baselineskip}{\bf (b)}}
 \includegraphics[height=6.525cm]{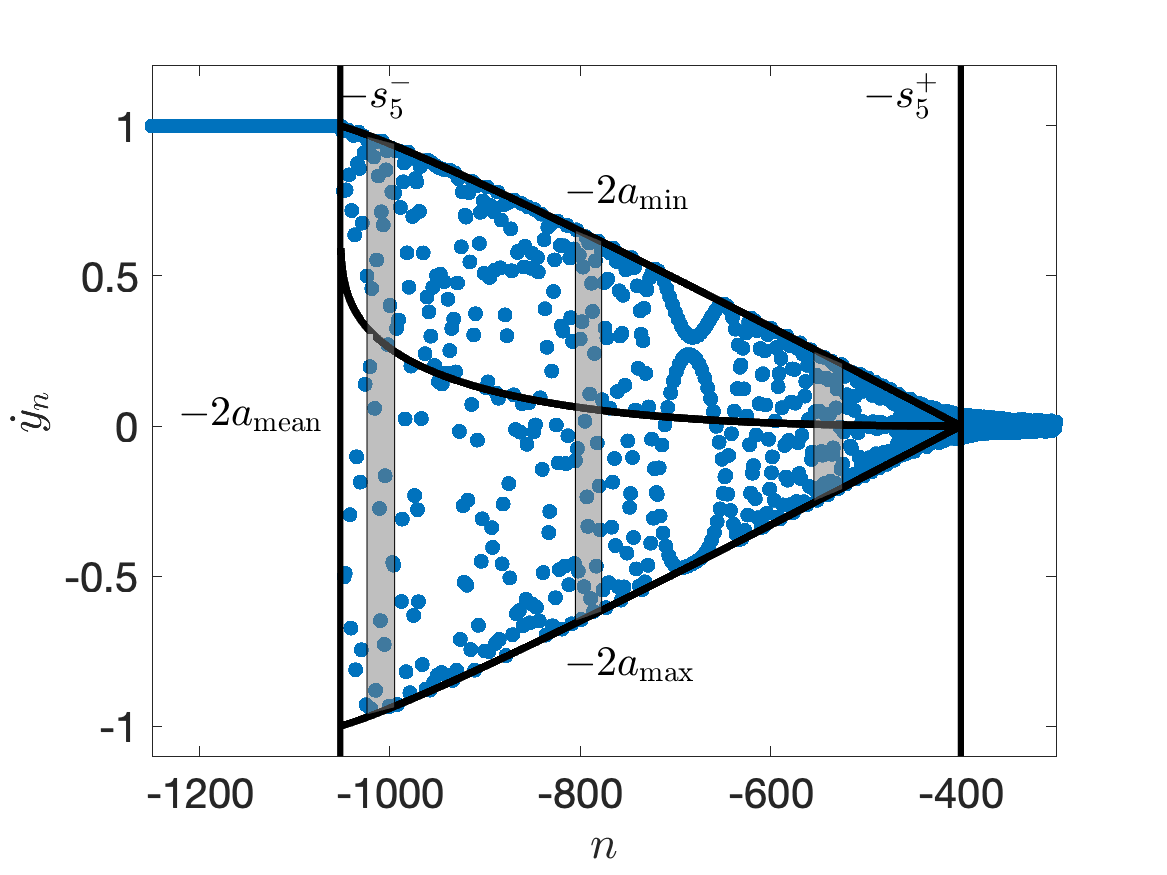} 
  \end{tabular}
  }
\centerline{
   \begin{tabular}{@{}p{0.33\linewidth}@{}p{0.33\linewidth}@{}p{0.33\linewidth}@{} }
     \rlap{\hspace*{5pt}\raisebox{\dimexpr\ht1-.1\baselineskip}{\bf (c)}}
 \includegraphics[height=4.3cm]{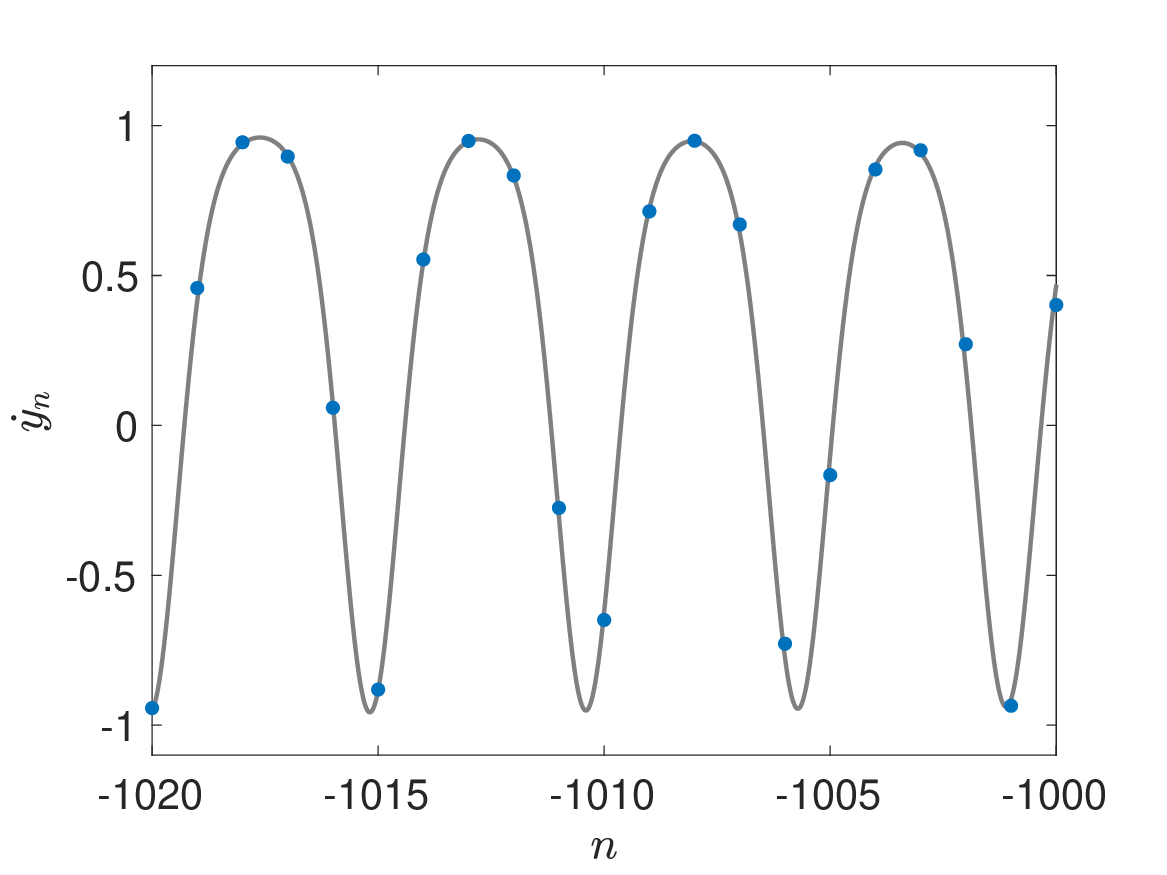} &
  \rlap{\hspace*{5pt}\raisebox{\dimexpr\ht1-.1\baselineskip}{\bf (d)}}
 \includegraphics[height=4.3cm]{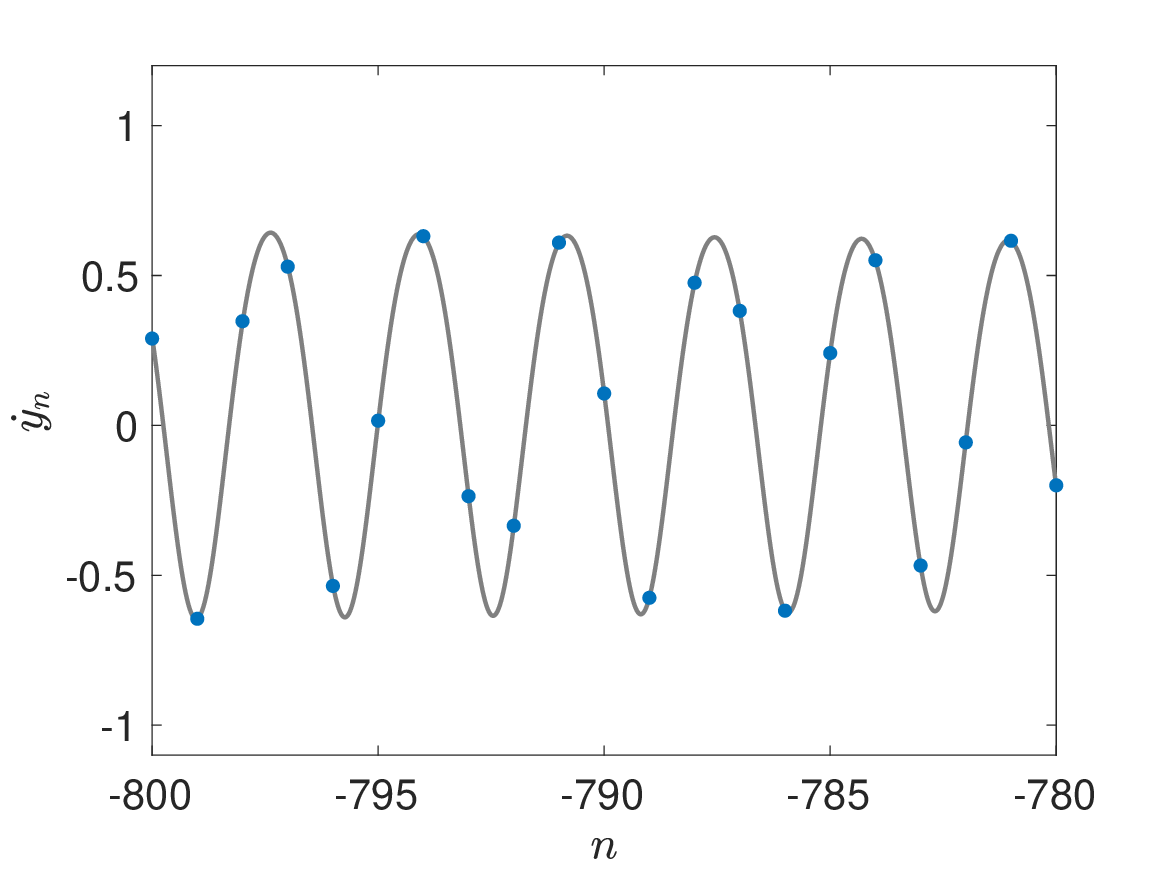} &
   \rlap{\hspace*{5pt}\raisebox{\dimexpr\ht1-.1\baselineskip}{\bf (e)}}
 \includegraphics[height=4.3cm]{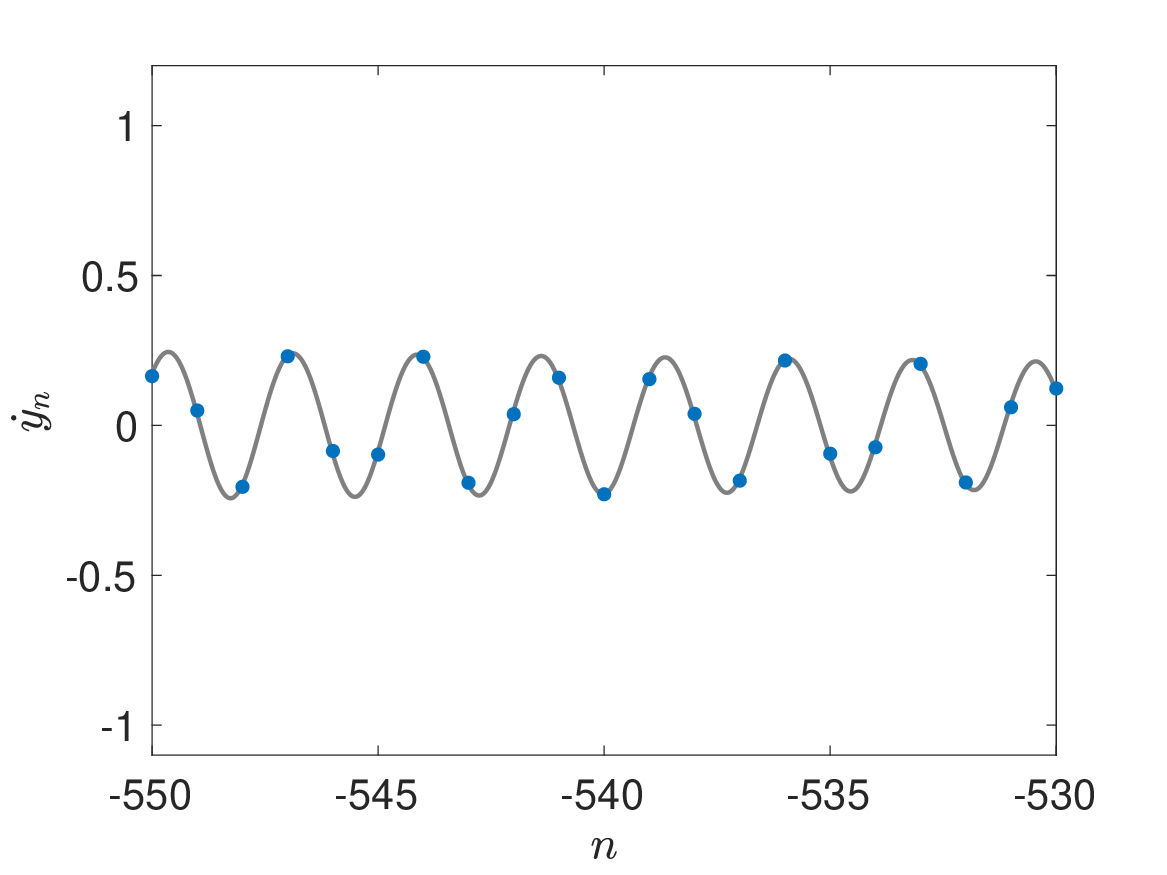} 
  \end{tabular}
  }
    \caption{Same as Fig.~\ref{fig:c1p5}, but with $c=0.5$. In panels (c) - (d) the corresponding windows
    of $m$ are $m\in[0.955,0.980]$, $m\in[0.705,0.742]$ and $m\in[0.280,0.340]$
    and the phase shift values  are $Z_0=2$, $Z_0=1.65$, and $Z_0=1.6$, respectively. 
    } 
    \label{fig:c0p5}
\end{figure}

 In terms of comparison, case 2 is very similar to case 1, with a few differences, which
 we highlight here. A 2D density plot
of a simulation initialized by Eq.~\eqref{e:TodashockIC} with $c=0.5$ and $N=8000$ nodes
is shown in Fig.~\ref{fig:c0p5}(a). 
In the figure, color intensity corresponds
to the velocity.   In the first region (where the lattice index
is negative) the leading and trailing edge of the DSW are predicted
by the $-s_5^-$ and $-s_5^+$ formulae, respectively, which
are defined in Eq.~\eqref{e:s5}.

%\paragraph{Step~1.}
In Fig.~\ref{fig:c0p5}(a),
a plot of $-n/s_5^-$ (solid white line) and $-n/s_5^+$ (dashed white
line) are shown, which compare very well to the leading and trailing
edge of the DSW in the first region. A similar plot of $-n/s_2^-$ (solid black line) and $-n/s_2^+$ (dashed black
line) are also shown, which compare very well to the leading and trailing
edge of the DSW in the second region. This comparison corresponds to
step 1 of Sec.~\ref{s:todadsw}.
These two regions correspond to
the genus-one regions of the Whitham equations. Unlike the simulation shown
for the $c>1$ case, the amplitude of the DSW decays to zero as the end
of the first Whitham zone is approached. Within the central region,
there are small amplitude linear waves that have amplitude that
decays proportionally to $t^{-1/3}$, which can be demonstrated  with the use of Fourier analysis \cite{MielkePatz2017}.
Note that, as usual, there is a discrepancy between the linear tails of the dispersive shock and the predictions of Whitham theory, which is due to the fact that one is considering a double limit in which both the wave amplitude and the parameter used in the Whitham expansion tend to zero; see \cite{Mark2016}.

%\paragraph{Step~3.}
In Fig.~\ref{fig:c0p5}(b), a zoom of the velocity profile of the DSW in the first Whitham zone 
at time $t=800$ is shown. The vertical solid lines are the theoretical predictions of the leading and
trailing edges located at the lattice positions $-800 s_5^-$ and $-800 s_5^+$, respectively. 
The minimum and maximum of the analytical prediction of the velocity
are given by Eq.~\eqref{e:bextreme} and the mean is given via \eqref{e:mean}. 
This comparison corresponds to step 3 of Sec.~\ref{s:todadsw}.
As in the $c > 1$ case, these formulae are parameterized by  $E_1,E_2,E_3,E_4$, 
whose definitions remain unchanged. For the present $0<c<1$ case, however,
the parameter $m$ belongs in the interval $m\in[0,1]$. The calculations
and subsequent comparisons are otherwise identical. We can also
compare the velocity profile itself to the theoretical
prediction, see Fig.~\ref{fig:c0p5}(c)-(d). This comparison 
corresponds to step 4 of Sec.~\ref{s:todadsw}.

We can also compare the time evolution against the theory.
Figure~\ref{fig:time_plots}(c,d) shows this comparison, and was generated in the same way
as Figure~\ref{fig:time_plots}(a,b), but for $c=0.5$.
In panel (b) the corresponding window of $m$ is $m\in[0.953,0.970]$
and the phase shift value is selected as $Z_0=1.53$.

\section{Conclusions and future challenges}
\label{s:conclusions}

In the present work, we have motivated, as well
as revisited the study of shock waves
%and rarefaction waves 
in the integrable realm of
the Toda lattice. We argued that the recent
development of numerous applications,
such as notably, recent studies of granular crystals and variants thereof~\cite{Molinari2009,HEC_DSW}, the
optical setting of coupled waveguides~\cite{fleischer2},
or that of tunable magnetic lattices~\cite{talcohen}
render this a timely and interesting topic of
exploration. Upon revisiting and homogenizing
the notation and fundamental solutions of
the Toda problem (including the solitonic and
periodic ones), we set up the Toda shock problem
in both its DSW and rarefaction variants for suitable
discrete jump initial data. We leveraged
the most general four-parameter family of elliptic
traveling waves of the Toda lattice~\cite{Teschl}
within the setup of Whitham modulation equations
of arbitrary genus for the Toda lattice,
based on the earlier key work of~\cite{blochkodama}.
We analyzed more concretely solutions of genus-zero and
genus-one, providing explicit expressions for the characteristic speeds of the latter. In addition
to illustrating the special harmonic and soliton limits
relevant to the two ends of the DSW, we showcased
how to use these elliptic solutions in conjunction
with the Whitham formulation to provide a full
characterization of the DSWs and rarefaction waves of
the Toda shock problem.

We believe that this first step in revisiting 
this time-honored problem may be an important
one. On the one hand, it suggests the relevance
of providing an analysis (from the integrable
and also from the modulation theory perspective)
of a wider class of Riemann problems, rather
than the arguably fundamental one examined herein.
In line with earlier studies concerning traveling
waves~\cite{Shen,epjp2020}, this effort may
pave the way for exploring regimes where 
the Toda problem may provide a controllably
suitable approximation to experimentally 
relevant settings such as granular crystals~\cite{Hascoet2000,Herbold07,Molinari2009},
hollow elliptic cylinder variants thereof~\cite{HEC_DSW}
or the recently explored tunable magnetic lattices~\cite{talcohen}.
It would also be interesting to use the soliton limit of the Whitham equations to study
the dynamics of solitons propagating on a slowly varying background, as in 
\cite{Hoefer2018,Hoefer2018a,NLTY2021v34p3583}.
Finally, it would be particularly interesting
to revisit other central integrable lattice
examples such as the well-known Ablowitz-Ladik
problem~\cite{AblowitzPrinariTrubatch}
and leverage their finite-genus solutions~\cite{Miller2,Teschl2,Vadim} in
order to solve the corresponding modulation 
equations~\cite{MILLER19961} and similarly 
describe the corresponding DSWs. This, in turn, could
be important towards both non-integrable (but
physically relevant) discrete
variants of the nonlinear Schr{\"o}dinger model.
%and towards ultra-discrete (in both space and time)
%ones.
Such directions constitute, in our view, an intriguing
program for future studies.

\bigskip
\noindent\textbf{Acknowledgments}
~\\[1ex]
The authors would like to thank the Isaac Newton Institute for Mathematical Sciences for support and hospitality during the programme Dispersive Hydrodynamics when work on this paper was undertaken (EPSRC Grant Number EP/R014604/1).
We would also like to thank Gennady El, Mark Hoefer, Yuji Kodama and Gerald Teschl for many insightful conversations on topics related to this work.
This work was partially supported by the National Science Foundation under grant numbers DMS-2009487 (GB) and DMS-2107945 (CC).

\small
\bibliographystyle{unsrt}
\bibliography{Chong}

\end{document}